\documentclass[reprint, amsmath, amssymb, aps, pre]{revtex4-2}
\usepackage{graphicx, dcolumn, bm, amsmath, amssymb, xcolor, enumitem, float}
\usepackage[spaces]{grffile}
\allowdisplaybreaks

\let\tempvec\vec
\renewcommand{\vec}[1]{\tempvec{{}#1}}
\let\tempbar\bar
\renewcommand{\bar}[1]{\tempbar{{}#1}}

\newcommand{\bmat}[2]{\def\arraystretch{#2}\begin{bmatrix}#1\end{bmatrix}}

\begin{document}
	\title{Thermal stability and secondary aggregation of self-limiting, geometrically-frustrated assemblies:  Chain assembly of incommensurate polybricks}
	\author{Michael Wang}
		\email{mwang@mail.pse.umass.edu}
	\author{Gregory Grason}
        \email{grason@umass.edu}
	\affiliation{Department of Polymer Science and Engineering, University of Massachusetts, Amherst, MA 01003}
	
	\begin{abstract}
		In geometrically frustrated assemblies, equilibrium self-limitation manifests in the form of a minimum in the free energy per subunit at a finite, multi-subunit size which results from the competition between the elastic costs of frustration within an assembly and the surface energy at its boundaries. Physical realizations -- from ill-fitting particle assemblies to self-twisting protein superstructures -- are capable of multiple mechanisms of escaping the cumulative costs of frustration, resulting in unlimited equilibrium assembly,  including elastic modes of ``shape-flattening'' and the formation of weak, defective bonds that screen intra-assembly stresses.  Here we study a model of 1D chain assembly of incommensurate ``polybricks'', and determine its equilibrium assembly as a function of temperature, concentration, degree of shape frustration, elasticity and inter-particle binding, notably focusing on how weakly cohesive, defective bonds give rise to strongly temperature-dependent assembly.  Complex assembly behavior derives from the competition between multiple distinct local minima in the free energy landscape, including self-limiting chains, weakly-bound aggregates of self-limiting chains, and strongly-bound, elastically defrustrated assemblies.  We show that this scenario, in general, gives rise to anomalous {\it multiple aggregation} behavior, in which disperse subunits (stable at low concentration and high temperature) first exhibit a primary aggregation transition to self-limiting chains (at intermediate concentration and temperature) which are ultimately unstable to condensation into unlimited assembly of finite-chains through weak binding beyond a secondary aggregation transition (at low temperature and high concentration).  We show that window of stable self-limitation is determined both by the elastic costs of frustration in the assembly as well as energetic and entropic features of inter-subunit binding. 
	\end{abstract}
 
	\maketitle

\section{\label{sec:introduction}Introduction}
  Geometric frustration arises when the local preferred ordering of a system cannot be realized on a global scale.  When this occurs, individual constituents of a system will often find complex ways of ordering on a large scale.  This concept has been studied in many contexts such as the frustrated ordering of magnetic spins on various lattices \cite{Wannier:1950,Vannimenus:1977,Bramwell:2001,Collins:1997}, nanoscale pattern formation on substrates \cite{Plass:2001,Lu:2003}, bent-core liquid crystals \cite{Reddy:2006,Takezoe:2006,Fernandez-Rico:2020,Niv:2018}, and the ordering of colloids on curved surfaces \cite{Meng:2014,Irvine:2010,Guerra:2018,Li:2019}.

		\begin{figure}
			\centering
			\includegraphics[width=\columnwidth]{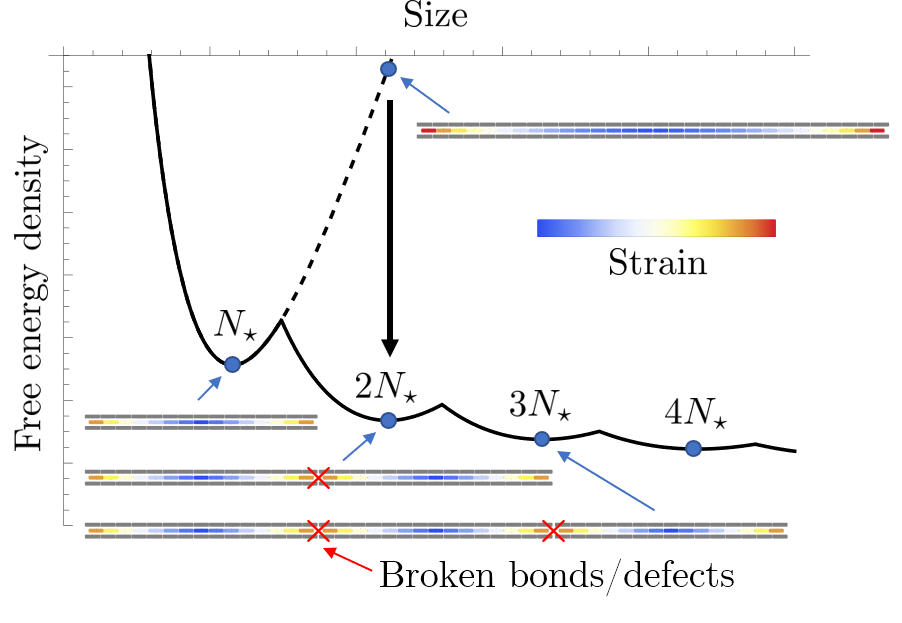}
			\caption{Attenuation of stress propagation through broken bonds and weak binding.  An assembly larger than its preferred size $N_{\star}$ can lower its free energy per subunit and relieve strain by breaking into multiple smaller, weakly-bound assemblies.}
			\label{fig:defect free energy}
		\end{figure}
        \begin{figure*}
            \centering
            \includegraphics[width=\textwidth]{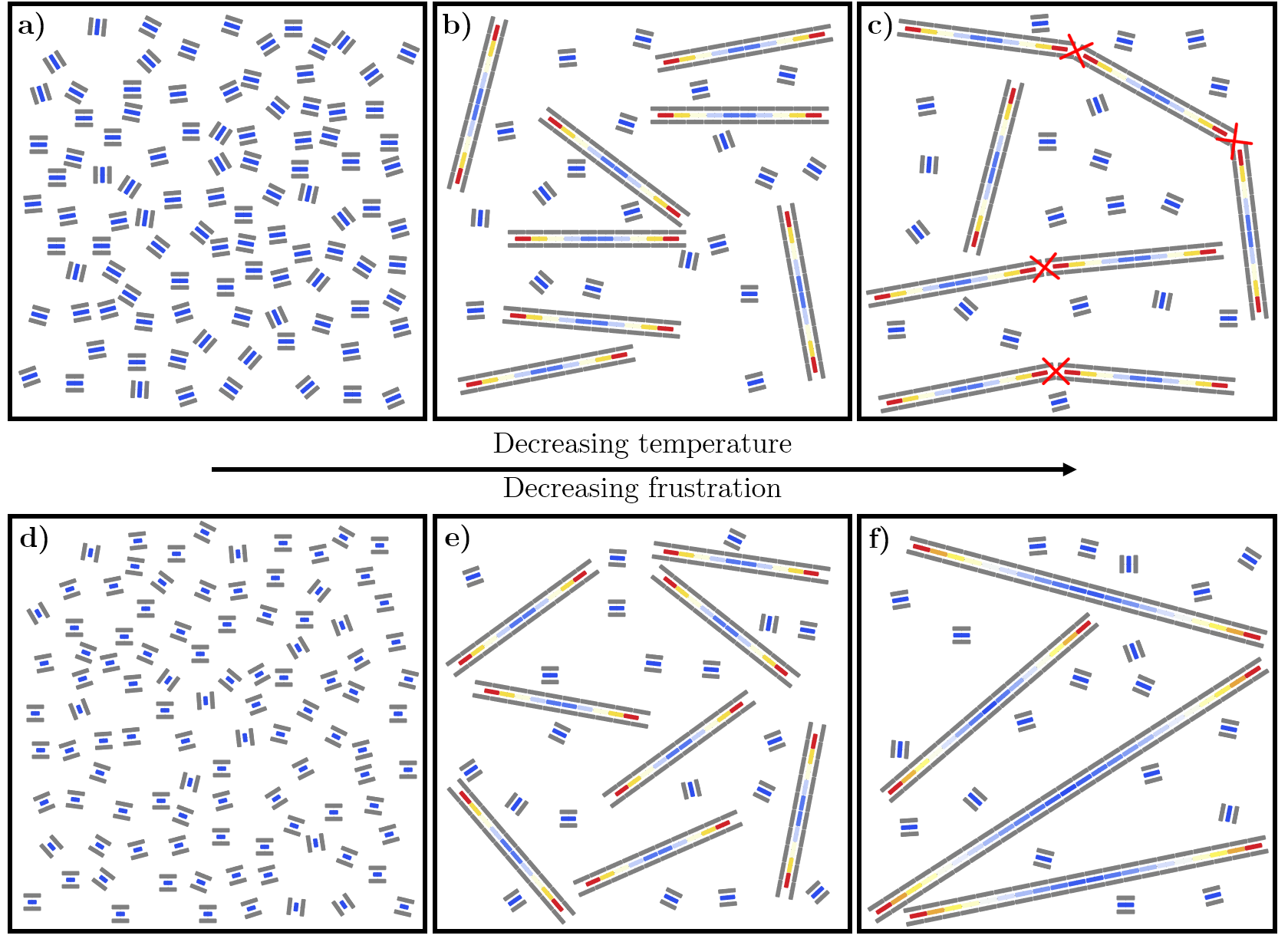}
            \caption{Example states of frustrated self-assembly.  Varying temperature results in \textbf{a)} dispersed subunits, \textbf{b)} self-limiting assemblies, and \textbf{c)} secondary aggregation of self-limiting assemblies through the formation of broken bonds or defects.  Varying frustration results in \textbf{d)} dispersed, highly-frustrated subunits, \textbf{e)} self-limiting assemblies, and \textbf{f)} shape-flattened, unlimited assemblies where frustration is not enough to overcome binding.}
            \label{fig:states}
        \end{figure*}
        
        Recently, there has been great interest in understanding the influence of geometric frustration on the morphologies as well as the sizes of self-assembled structures of misfitting building blocks~\cite{Grason:2016,Lenz:2017,Meiri:2021}.  The concept of {\it geometrically frustrated assembly} (GFA) has been applied to a range of existing systems in soft matter, from protein bundles \cite{Aggeli:2001, Turner:2003, Grason:2007, Yang:2010, Brown:2014, Hall:2016} and chiral membranes of liquid crystals and surfactants \cite{Ghafouri:2005, Armon:2014, Zhang:2019, Matsumoto:2009, Gibaud:2012,Sharma:2014} to more recent efforts to design and assemble intentionally misfitting particles \cite{Berengut:2020,Uchida:2017,Serafin:2021,Lenz:2017,Tyukodi:2022,Tanjeem:2022,Spivack:2022,Hall:2023}.  When such building blocks assemble through an attractive interaction, the resulting structure, as well as the blocks themselves and their interactions, must deform as the assembly grows.  This competition between the attractions of the building blocks and the build up of strain regulates the equilibrium sizes and morphologies of the resulting assemblies, a phenomenon known as self-limiting assembly \cite{Grason:2016,Hagan:2021}.  Basic questions regarding the fundamental and practical importance of GFAs include how does the self-limiting size depend on shape and interactions of given building blocks, and how large can the self-limiting size be relative to that subunit size?

        Self-limitation in GFA relies on the propagation of intra-assembly gradients of local strain, whose effects lead to the accumulation of elastic costs that grow super-extensively with size~\cite{Meiri:2021}. Mechanisms that relax or obstruct the propagation of elastic effects to large sizes and thus limit the maximal range of self-limiting sizes, can be broadly grouped into two categories~\cite{Grason:2016}.  The first, dubbed as ``shape flattening'', occurs when the shape of misfitting particles is sufficiently soft that it is overwhelmed by the strong, inter-particle cohesive binding, deforming the particle assembly into uniform, unfrustrated motifs, and giving way to unlimited (i.e. bulk) assembly.  Such ``soft'' mechanisms of frustration escape are controlled by elastic parameters that control both shape deformation and intra-particle strains.  Alternatively, the second category of frustration escape mechanisms relies on inelastic defects in the assembly that, at least partially, mitigate the propagation of frustration induced stress.  For 2D ordered assemblies, these may take the form of topological defects, e.g. disclinations and dislocations, that screen frustration \cite{Irvine:2010,Azadi:2014,Guerra:2018,Li:2019,Grason:2010,Bruss:2013, Hackney:2023}.  More generally, a defect may take the form of any partial bond that localizes frustration induced stress, such as an internal crack or weak-bond, but nevertheless acquires some cohesion (see e.g. examples in assembly models of hyperbolic tubules in ~\cite{Tyukodi:2022,Hall:2023}).  
        
        In this article, we study a 1D model of GFA capable of exhibiting both modes of frustration escape, and in particular, aim to understand the generic effects of ``weak'', defective binding on the thermal stability of self-limiting assembly.  Our model is inspired by the so-called DNA origami based ``polybrick'' particle, developed and studied by Berengut and coworkers \cite{Berengut:2020}, shown to exhibit self-limiting chain assembly.  Figure \ref{fig:defect free energy} illustrates schematically the generic effects of weak binding in a 1D frustrated assembly on the free energy per subunit landscape.  While the accumulation of frustration stresses between {\it strongly-bound} particles leads to a primary minimum at a well-defined size $N_*$, it is straightforward to see that weak-binding between two or more strongly-bound chains can lead to additional local minima in the free energy per subunit at integer multiples of the primary aggregate size.  At zero-temperature, any weak cohesive interaction guarantees that the ground state is an infinite chain of weakly-bound aggregates, i.e. an unlimited aggregate.  At finite temperature, secondary assembly of primary aggregates is controlled by the free energy of weak-binding, which may be favorable or unfavorable depending on both the energy gain and entropy cost of weak binding.  In general, this illustrates that geometrically frustrated self-assembly is characterized by multiple local minimal at distinct sizes which compete for stability.  Based on the exact solution of the ideal 1D assembly behavior of the frustrated polybricks, we illustrate two basic and generic consequences of weak, defective binding in frustrated assembly.  First, we show that self-limitation is only possible above a minimal finite temperature.  Second, we show that the competition between the primary aggregation into strongly-bound assemblies and the weak binding of those assembled chains leads to an exotic type of {\it secondary aggregation} behavior.  
        
        Canonical assembly, like the spherical micellization of surfactants, is marked by a single pseudo-critical aggregation transition between a dispersed state at low concentration and a micelle-dominated state at high concentration~\cite{Israelachvili:2011,Israelachvili:1976}.  Cases where there are multiple, nearly-degenerate local minima in the free energy per subunit exhibit more complex scenarios~\cite{Hagan:2021}, where there is a primary aggregation transition from a dispersed state into a finite aggregate state, followed by a {\it secondary aggregation transition} from primary aggregates to a secondary state of aggregates (of typically higher aggregate mass) at further higher concentration or lower temperature.  For surfactants, such secondary micellization transitions have been studied in the context of concentration-dependent transitions from spherical to cylindrical micelles, where the nature of the transition is controlled by microscopic energetics of molecular packing in sphero-cylindrical aggregates \cite{Porte:1984, May:2001, Bernheim:2000, Magnus:2016}.  

        Here, we show that the secondary aggregation behavior is a generic feature of GFAs due to unavoidable possibility of weak, partial binding between attractive subunits.  In what follows, we introduce an exactly solvable model of a frustrated polybrick assembly, and predict the assembly behavior as a function of {\it frustration}, {\it concentration}, and {\it temperature}, showing in general that self-limitation is possible at intermediate regimes of these basic parameters, shown schematically in Fig. 2.  As temperature is decreased (or concentration is increased), we predict a sequence of states from dispersed monomers to self-limiting to defective and unlimited (i.e. weakly-bound chains of chains) assembly.  As frustration is decreased (below a critical temperature), we predict a sequence of states from dispersed monomers to self-limiting to strongly-bound, unlimited chains.  We relate these stability windows to microscopic parameters that control intra-assembly elasticity, as well as entropy and energy of binding.  The manuscript is structure as follows.  In Section \ref{sec:model}, we introduce our model of an incommensurate chain of ``polybricks'' and the finite-temperature corrections to the free energy per subunit.  In Section \ref{sec:criteria}, we describe the ideal aggregation of linear chains and establish criteria for determining whether a system is dispersed, self-limiting, or defective.  In Section \ref{sec:phase diagrams}, we describe the distinct free energy per subunit landscapes possible in our model, their corresponding concentration-dependent aggregation, and summarize the equilibrium assembly behavior of frustrated polybricks in the temperature-frustration plane.

	\section{\label{sec:model}Frustrated, incommensurate chain model}
		\begin{figure*}
			\centering
			\includegraphics[width=\textwidth]{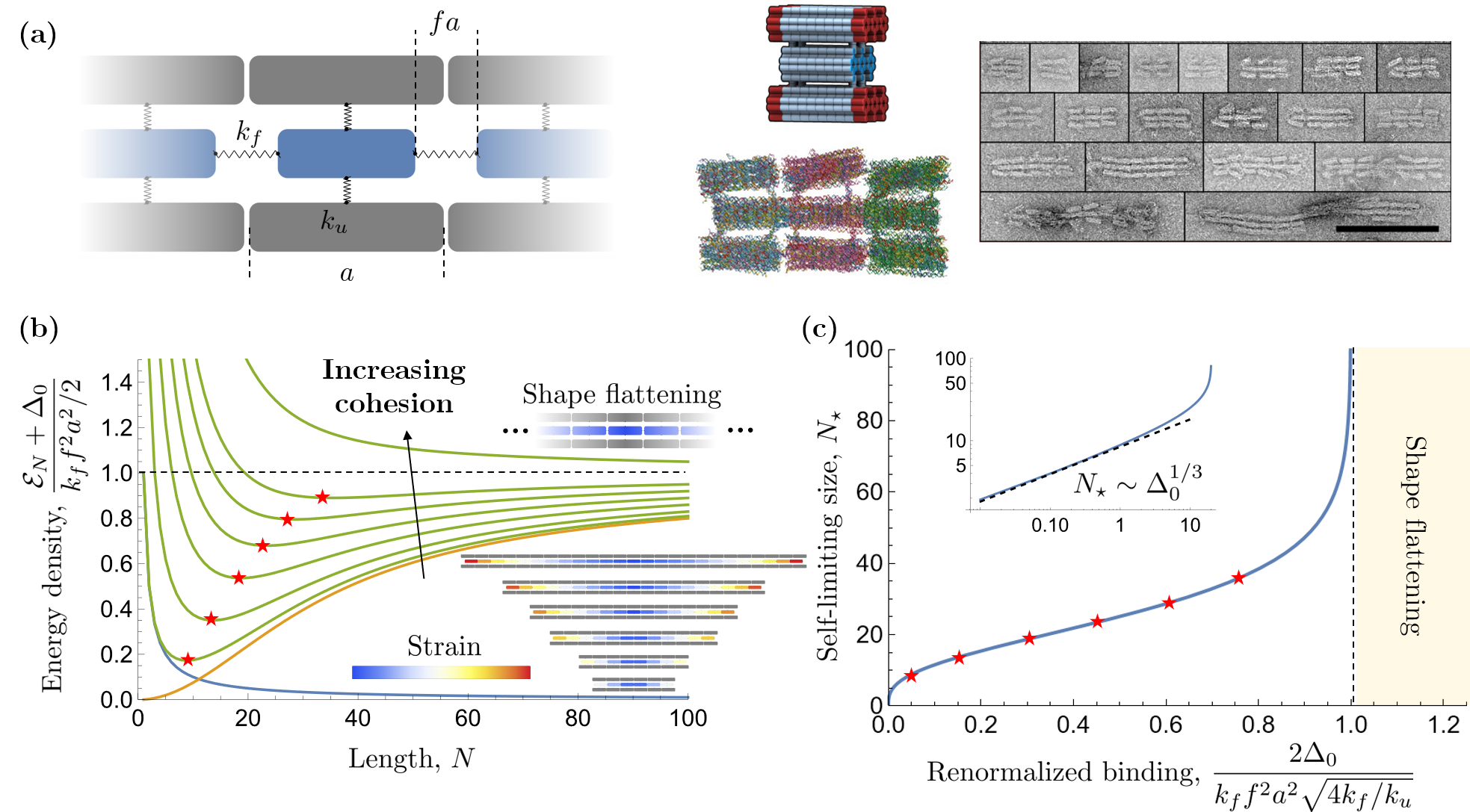}
			\caption{\textbf{a)} Incommensurate chain model (left) inspired by experiments \cite{Berengut:2020} (right).  \textbf{b)} Energy per subunit of linear chains of length $N$ for increasing cohesion (binding energy $\Delta_0$) with ratio of elasticities $k_f/k_u=100$ ($k_f=100$, $k_u=1$) and frustration $fa=0.05$.  Red stars indicate the local energy minima and the corresponding structures and their strains are shown.  \textbf{c)} Self-limiting size as a function of the cohesion between subunits.}
			\label{fig:model}
		\end{figure*}
	
		To gain insight into the role of temperature on the self-limiting assembly of GFAs, we start by introducing a simple solvable model of a linear chain of incommensurate subunits inspired by recent experiments of DNA origami, incommensurate ``polybrick'' particles \cite{Berengut:2020}.  We begin by defining the (zero-temperature) energetics of strongly-bound chain assembly in the model

  \subsection{Energetics of strongly-bound, frustrated chains}
  
  The subunits in this system consist of three blocks (i.e. rectangular prisms of a honeycomb DNA lattice): a short block of length $(1-f)a$ sandwiched between two longer blocks of length $a$, as shown in Figure $\ref{fig:model}$a.  Here the quantity $f$ characterizes the amount of shape mismatch, namely the {\it frustration}, in a linear assembly of such subunits with the central blocks preferring a lattice spacing of $(1-f)a$ that is incommensurate with the preferred lattice spacing $a$ of the outer blocks.  Within a subunit, our model considers the central block to be elastically attached to the two outer blocks with a spring of stiffness $k_u$ favoring the middle block to be centered between the outer two longer blocks and penalizing lateral (i.e. sliding) displacements $u$ relative to the centered position.  Hence, $k_u$ parameterizes the intra-subunit deformablility.  To keep the analysis simple, we assume that the two outer blocks move together and act as a backbone along which the central block can slide, although it is possible for a more complex array of intra-particle deformation modes to renormalize the effective parameters described here.  Finally, when two subunits are strongly bonded, the central blocks elastically interact through a spring of stiffness $k_f$.  That is, binding favors a specific face-to-face contact, but the incommensurate geometry of blocks in general requires some strain displacement from that contact.  In the case of DNA origami polybricks, the inter-subunit stiffness can derive from the deformation of bound single-stranded DNA bases that mediate attractions between the particle faces in combination with the compression of the repulsive brushes that coat  flanking blocks.  More generally, the stiffness of specific interactions between frustrated particles can be manipulated via surface functionalization with complementary single-stranded DNA sequences \cite{Cui:2022,Rogers:2011}.
		
		The energy of a linear chain of $N$ {\it strongly-bound} subunits is
		\begin{align}
			\begin{split}
				E_N[\{u_n\}]=&-(N-1)\Delta_0+\sum_{n=1}^N\frac{1}{2}k_uu_n^2\\
				&+\sum_{n=1}^{N-1}\frac{1}{2}k_f\left(u_{n+1}-u_n+fa\right)^2,
			\end{split}
		\end{align}
		where $u_n$ is the position of the central block relative to the outer blocks of the $n$th subunit.  The first term describes the effect of cohesive gain of strong bonds in the system, where $\Delta_0$ is the binding energy of the ideal, unstrained bond.  The second and third terms correspond to the intra-subunit deformation energy, and the inter-subunit bond-stretching energy, respectively.  This energy can be minimized with respect to the displacements $\{u_n\}$, the details of which we leave for Appendix \ref{app:exact solution}.  The per-subunit ground state energy is
		\begin{equation}
			\mathcal{E}_N=\frac{E_N}{N}=-\left(1-\frac{1}{N}\right)\Delta_0+\mathcal{E}_N^{(\rm ex)},
			\label{eq:ground state energy}
		\end{equation}
		where $\mathcal{E}_N^{(\rm ex)}$ is the {\it excess energy} per subunit built up from the accumulation of strain due to the incommensurate lengths of the blocks.  The per-subunit ground state energy is shown in Figure \ref{fig:model}b.  The excess energy density, indicated by the orange curve, has a useful continuum approximation given by
		\begin{equation}
			\mathcal{E}_N^{(ex)}\approx\frac{1}{2}k_ff^2a^2\left(1-\frac{\tanh N/\sqrt{4k_f/k_u}}{N/\sqrt{4k_f/k_u}}\right).
		\end{equation}
		The quantity $\sqrt{4k_f/k_u}$ can be interpreted as the size scale over which strains can accumulate.  It is also the ratio of the intra-subunit deformability to the inter-subunit deformability.  For small assemblies $N\ll\sqrt{4k_f/k_u}$, the per-subunit excess energy grows superextensively as $\mathcal{E}_N^{(\rm ex)}\sim k_uf^2a^2N^2$ as the assembly strains more and more to bind new subunits, i.e. in this regime the central brick must displace more and more $u_n \sim f a N$ to account for the accumulating length mismatch between tightly bound units .  At large assembly sizes $N\gg N_{\rm flat} = \sqrt{4k_f/k_u}$, the subunits can only internally deform so much before it is favorable to strain the bonds between them instead, i.e. they begin to adopted uniformly strained interactions (with the exception of a boundary layer) and the energy density plateaus to $\mathcal{E}_{\infty}^{(\rm ex)}=k_ff^2a^2/2$.  As shown in the example assemblies in Figure \ref{fig:model}b, a majority of the strain-gradient  is expelled to the boundaries while the bulk subunits remain relatively uniformly strained.  This is known as \textit{shape flattening}.  The frustration and strain accumulation, whose per particle cost grows as $\mathcal{E}_{\rm strain}(N) \approx k_u (a f N)^2$ for small sizes, competes with the binding of new subunits as the binding energy density, indicated by the blue curve, decreases as $\sim\Delta_0/N$ and favors larger assemblies.  This competition leads to an energetically favorable, self-limiting size $N_{\star}(T=0)\sim\left(\Delta_0/k_uf^2a^2\right)^{1/3}$, as shown in Figure \ref{fig:model}c.  As one increases the strength of binding or decreases the frustration, the self-limiting size increases.  However, above a certain threshold binding strength (or below a threshold frustration),assemblies enter the shape flattening regime, i.e. $N_* \gtrsim N_{\rm flat}$ where the accumulative cost of sliding strain exceeds the cost to deform the inter-particle blocks uniformly.  As binding becomes more cohesive, or particle shapes become less frustrated, the energy per subunit no longer has a minimum at finite $N$, and the ground state structure becomes unlimited in size.  This condition can be estimated by equating accumulating intra-particle shear energy to the cost of uniform bond strain.  From this condition, we see that shape-flattening limits the range of equilibrium self-limitation to an {\it upper limit} on the range of finite-size of chains which occurs at a corresponding {\it minimal value} of frustration (see Appendix \ref{subapp:continuum})
		\begin{equation}
		    N_* \lesssim N_{\rm flat}\approx \sqrt{k_f/k_u}; \ f \gtrsim f_{\rm flat} (T=0)\approx \Delta_0^{1/2} k_f^{-3/4}k_u^{1/4},
        \end{equation}
	    where here we consider the case of fixed cohesive energy per strong bond ($\Delta_0$). In general, these conditions suggest that increasing the value of {\it interaction stiffness} $k_f$ relative to the intra-particle stiffness $k_u$ favors larger range of self-limiting sizes.
     
		\subsection{\label{subsec:bond entropies}Vibrational and orientational contributions to free energy}
			\begin{figure*}
				\centering
				\includegraphics[width=\textwidth]{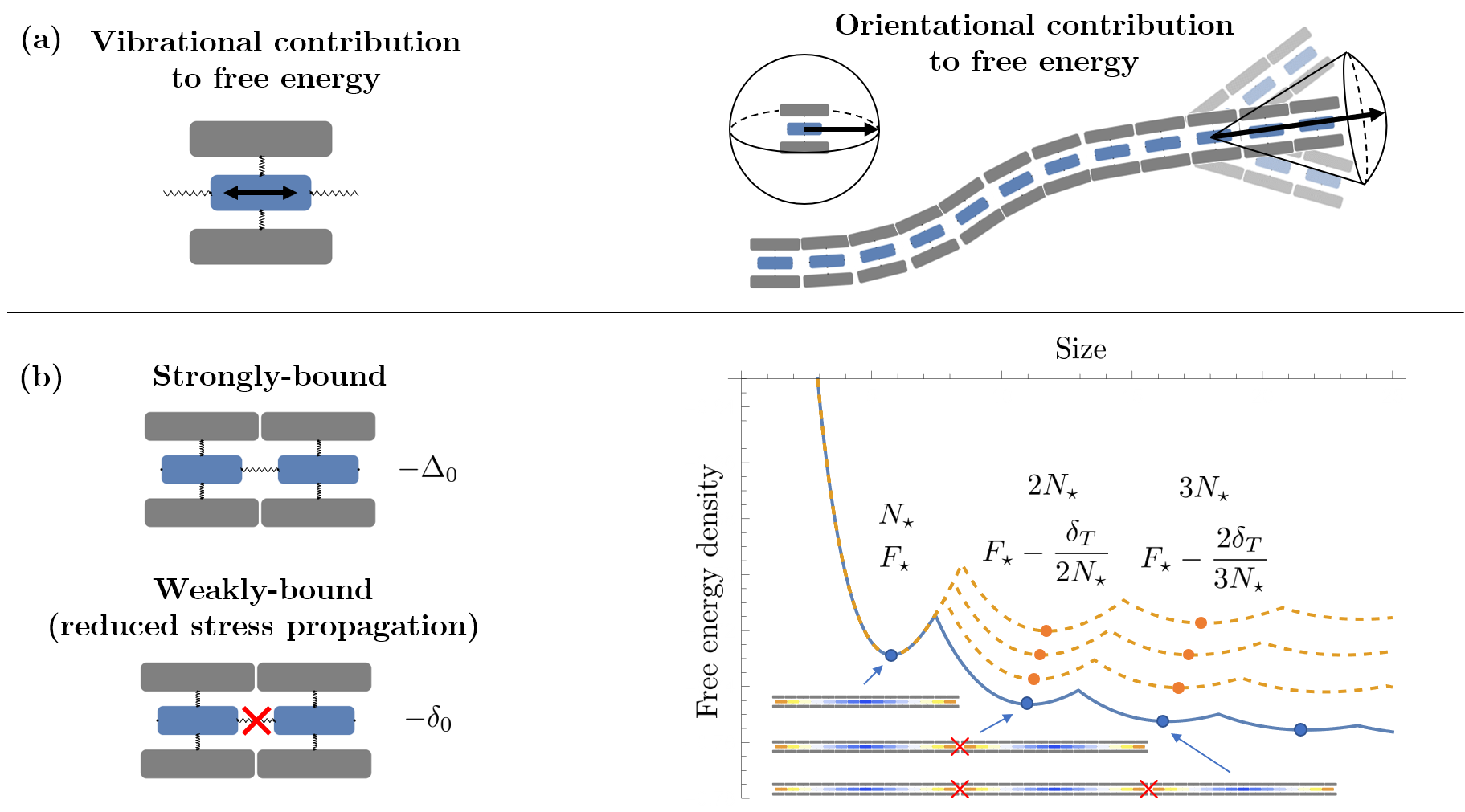}
				\caption{\textbf{a)} Bound subunits are more restricted in their vibrational and orientational modes, leading to a reduction of vibrational and orientational entropies.  \textbf{b)} Strongly-bound subunits have a binding energy $-\Delta_0$ while weakly-bound subunits, which results from broken bonds or any residue binding, have a binding energy $-\delta_0$.  The free energy landscape contains infinitely many local minima corresponding to the primary self-limiting assembly and defective aggregates of weakly-bound self-limiting assemblies.  Changing the temperature-dependent weak-binding free energy $\delta_T$ (see Section \ref{subsec:landscape}) changes whether the defective aggregates are favorable $-\delta_T<0$ or unfavorable $-\delta_T>0$.}
				\label{fig:bond entropies}
			\end{figure*}
		
			     At finite temperature, the vibrational and orientational degrees of freedom that describe conformational flucutations of aggregates lead to associated entropic contributions to the free energy of assembly.  A free subunit can access all possible orientations.  However, when strongly bound to another subunit, its possible orientations relative to its bound neighbor are reduced.  We model this by restricting a strongly-bound unit to a cone with angle $\theta_{\Delta}$ (Figure \ref{fig:bond entropies}a).  This leads to a reduction in the orientational entropy, which can be computed from the partition function
			\begin{equation}
				Z_{\rm orient}=\left(\int_{0}^{2\pi}\int_{0}^{\theta_{\Delta}}\frac{d\Omega}{4\pi}\right)^{N-1}.
			\end{equation}
			The vibrational contribution can be computed by considering collective sliding modes in a chain of $N$ strongly-bound subuunits (i.e. phonons) $\{\delta u_n\}$ about the ground state positions of the central blocks.  The partition function is
			\begin{equation}
				Z_{\rm vib}=\int\left(\prod_{n=1}^{N}d\delta u_n\right)e^{-\beta E_N[\{u_n^{(\rm ex)}+\delta u_n\}]}.
			\end{equation}
			We leave the details in Appendix \ref{subapp:entropy}.  The free energy per subunit of the linear chain can be computed as 
			\begin{equation}
				\mathcal{F}_N=-\frac{k_B T}{N}\ln Z_{\rm orient}Z_{\rm vib}=\mathcal{E}_N+\delta\mathcal{F}_N^{(\rm orient)}+\delta\mathcal{F}_N^{(\rm vib)},
			\end{equation}
			where $\mathcal{E}_N$ is the per-subunit ground state energy (Eq. (\ref{eq:ground state energy})) and $\delta\mathcal{F}_N^{(\rm orient)}$ and $\delta\mathcal{F}_N^{(\rm vib)}$ are the orientational and vibrational contributions to the free energy given by
			\begin{subequations}
				\begin{align}
					\delta\mathcal{F}_N^{(\rm orient)}&=\left(1-\frac{1}{N}\right)k_B T\ln\frac{2}{1-\cos\theta_{\Delta}},\\
					\delta\mathcal{F}_N^{(\rm vib)}&\approx\left(1-\frac{1}{N}\right)k_B T\ln\left[\frac{1}{2}\left(1+\sqrt{1+\frac{4k_f}{k_u}}\right)\right].
				\end{align}
			\end{subequations}
			Observe that the forms of these contributions allow us to rewrite the total per-subunit free energy as
			\begin{equation}
				\mathcal{F}_N=-\left(1-\frac{1}{N}\right)\Delta_T+\mathcal{E}_N^{(\rm ex)},
			\end{equation}
			where $\Delta_T$ is the temperature-corrected binding {\it free energy} takes on the form
			\begin{equation}
				\Delta_T=\Delta_0-Ts_{\Delta}.
			\end{equation}
			Notably, the quantity $s_{\Delta}>0$ captures the effect of temperature on binding.  In essence, the vibrational and orientational entropy costs of binding effectively weaken the energetic gain of assembling.  Note that $T_{\Delta}=\Delta_0/s_{\Delta}$ is the temperature above which (unfrustrated) bonds are melted and the energetically favorable state is a single subunit for $f=0$.  To get a sense for the value of $s_{\Delta}$, suppose that $\theta_{\Delta}\sim1^{\circ}-10^{\circ}$ and $k_f/k_u\sim1-100$, which gives $s_{\Delta}\sim(5-10)k_B$.  Note that if the full binding energy is of order $10k_BT$, then the vibrational/orientational melting of the bond occurs on order few $k_BT$.
		
		\subsection{\label{subsec:landscape}Free energy landscape of weak binding}
			As the size of a frustrated assembly grows, it will often find ways of escaping and relieving frustration through the formation of defects that attenuate or stop the the propagation of stresses \cite{Grason:2016}.  In our case, we represent this attenuation of stress propagation as broken bonds between the central block of two neighboring subunits (Figure \ref{fig:bond entropies}b), which we call weak, or ``defective", bonds.  In such a state, we assume that there can still be some weaker, residual cohesion between subunits that has a characteristic binding energy $-\delta_0\geq -\Delta_0$.  In a manner similar to the case of fully bound subunits (see Appendix \ref{subapp:entropy}), the entropic contributions to the weak-binding results in a temperature-corrected weak-binding \textit{free energy}
			\begin{equation}
				\delta_T=\delta_0-Ts_{\delta}.
			\end{equation}
            As with strong bonds, finite temperature has the effect of making weak binding, or defects, less favorable due to the entropic costs of reduced rotational freedom of binding together two otherwise freely-rotating chains.   Hence, above a temperature $T_{\delta}=\delta_0/s_{\delta}$, they become entropically unfavorable since $-\delta_T>0$.  As we illustrate below, the entropic cost of defect formation has a critical role on the thermodynamic stability of self-limiting states.
			
			To construct the free energy landscape of the incommensurate chain model including both strong (stress-propagating) and weak (defective, stress-attenuating) states of sub-unit binding, we start with the free energy $\mathcal{F}_N$ of a linear chain of $N$ strongly bound subunits.  As the size of the chain grows from a single subunit, the free energy initially decreases due to binding until the chain reaches the self-limiting size (a local minimum at primary aggregate size $N=N_*$) after which it increases due to stress accumulation.  Eventually, when enough stress builds up, it becomes energetically favorable for the linear chain to break into two weakly-bound pieces, the per-subunit free energy of which can be approximated as $\mathcal{F}_{N/2}-\delta_T/N$, where the first term is the per-subunit free energy of each half and the second term is the weak-binding free energy distributed over the $N$ subunits.  This reasoning can be continued for a linear chain broken up into $M\le N$ weakly-bound pieces with an approximate per-subunit free energy $\mathcal{F}_{N/M}-(M-1)\delta_T/N$.  Finally, for each size $N$, there will be some number $M\ge1$ of strongly-bound subchains, held together by $M-1$ weak bonds, that is the thermodynamically hierachical chain to form.  Thus, the free energy can be taken to be
			\begin{equation}
				\mathcal{F}_N=\min_{1\le M\le N}\left\{\mathcal{F}_{N/M}-\frac{(M-1)\delta_T}{N}\right\}.
			\end{equation}
			The key result of the hierarchy of strong and weak bonds is shown in Figure \ref{fig:bond entropies}b right, illustrated for series increasing values weak bond free energies.  In general, if $N_{\star}$ is the self-limiting size, then the subsequent local minima (separated by barriers) of the free energy are roughly $2N_{\star},3N_{\star},\dots$, corresponding to multiple self-limiting assemblies weakly aggregating together.  This serves as the starting point for a useful approximation we discuss in Appendix \ref{app:multiple minima approximation}.  The value of $-\delta_T$ regulates which of these local minima is the global one.  When $-\delta_T>0$ the entropy cost of defects leads to a global ground-state at the primary minimum $N=N_*$, a single strongly bound self-limiting aggregate, defined by the balance between frustration and cohesion.  When $-\delta_T<0$ weak-binding of strong-aggregates is favored, and the global ground state transitions to infinite, defective chains of primary aggregates.  In this context, it is clear that as $T\to0$ {\it any} weak cohesion in defective bonds leads to an unlimited chain of primary aggregates as the ground-state.

	\section{\label{sec:criteria}Ideal aggregation and self-limiting vs. unlimited assembly criteria}
		\begin{figure*}
			\centering
			\includegraphics[width=\textwidth]{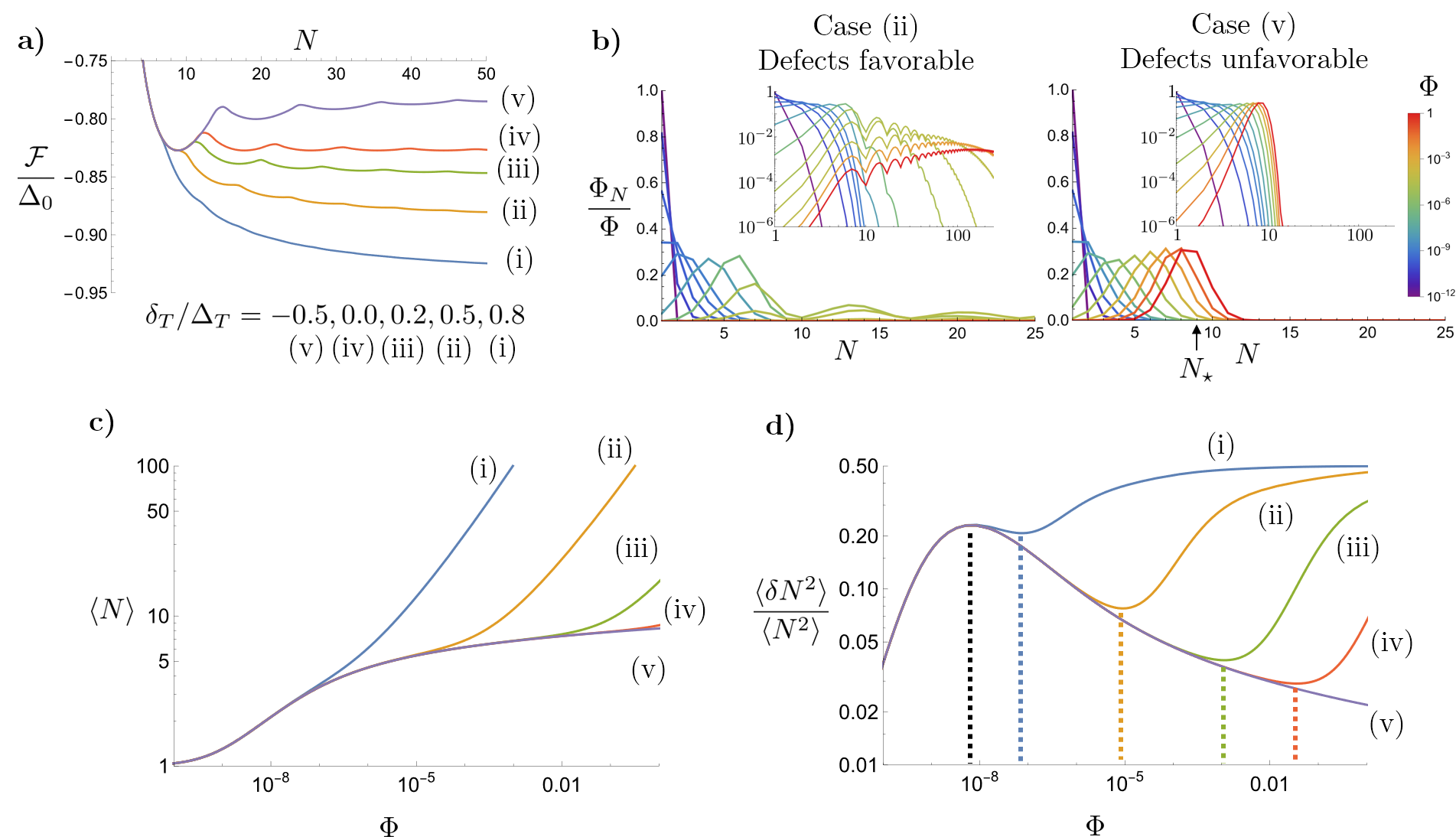}
			\caption{Role of varying the weak-binding free energy.  \textbf{a)} Example free energies for favorable (i, ii, iii), neutral (iv), and unfavorable (v) weak-binding for $k_f/k_u=100$ and $k_B T/\Delta_0=0.05$.  The self-limiting size is fixed at $N_{\star}\sim 9$.  \textbf{b)} Mass distributions of subunits for favorable weak-binding (left), which leads to unlimited growth and multiple aggregation of self-limiting assemblies, and unfavorable weak-binding, which leads to the stabilization of self-limiting assemblies.  Insets show mass distributions and sizes on log scales.  \textbf{c)} Average assembly size.  \textbf{d)}  Relative size fluctuations.  The black dotted line corresponds to the local maximum or the onset of self-limiting assembly while the colored dotted lines correspond to local minima or the onset of weakly-bound aggregation.}
			\label{fig:example}
		\end{figure*}
		Beyond the energetics and entropy of internal configurations of assembled chains, equilibrium assembly also depends on the translational entropy associated with distributing subunits among different populations of aggregates at fixed temperature and concentration.  We do this by applying ideal aggregation theory to a system with a fixed concentration of subunits \cite{Israelachvili:2011, Hagan:2021}, which assumes that concentrations are low enough that interactions between different aggregates have a negligible effect on their free energy.  Let $\Phi_N$ be the volume fraction of subunits in assemblies of size $N$.  The total free energy of the system is
		\begin{equation}
			\mathcal{F}_{\rm tot}=\sum_{N=1}^{\infty}\Phi_N\left(\mathcal{F}_N+\frac{k_B T}{N}\ln\frac{\Phi_N}{Ne}\right),
		\end{equation}
		where the logarithmic term represents the translational entropy of $N$ subunits sharing a single center of mass degree of freedom.  Minimizing the total free energy with respect to the volume fraction, $\Phi$, yields the {\it law of mass action}, $\Phi_N=N\left(\Phi_1e^{-\beta\mathcal{F}_N}\right)^N$.  The dispersed subunit volume fraction $\Phi_1$ can be determined from mass conservation, that is, the total volume fraction of subunits must remain fixed or
		\begin{equation}
			\Phi=\sum_{N=1}^{\infty}N\left(\Phi_1e^{-\beta\mathcal{F}_N}\right)^N.
		\end{equation}
		This relation is an equation of state that relates the mass of subunits in distinct aggregate populations to the total mass of subunits, requiring that all aggregates are in {\it chemical equilibrium}.  Here, the concentration of free monomers parameterizes the chemical potential of subunits in the system, $k_BT\ln\Phi_1$ \cite{Israelachvili:2011}.
		
		To gain intuition into the role of the weak binding free energy $\delta_T$, we consider a simple example shown in Figure \ref{fig:example} where we vary only the weak binding free energy while keeping the self-limiting size, excess elastic energy of frustration and temperature fixed.  When the weak-binding free energy $\delta_T$ is comparable to the full binding free energy $\Delta_T$, the energy landscape loses the self-limiting minimum (Figure \ref{fig:example}a, cases (i) and (ii)) as it is thermodynamically favorable to break up strongly-bound chains into smaller pieces, which do not propagate frustration induced stress.  As one lowers the weak binding free energy, the self-limiting minimum as well as the local minima for weakly-bound aggregates reappear (cases (iii-v)), leading to corrugations in ${\cal F}(N)$ associated with states of integer numbers of primary aggregate chains held together by weak, defective bonds.  In particular, depending on the entropic costs of weak binding, it is possible for the weak binding free energy to change signs, leading to weakly bound aggregates whose free energies are higher than that of the self-limiting state (case (v)).  
		
		The temperature-dependent sign change in the weak-binding free energy plays a crucial role in the stabilization of self-limiting assemblies, which can be seen by examining the subunit mass distributions for two examples, shown in Figures \ref{fig:example}b.  At low concentrations, most of the subunits remain fairly dispersed as free monomers due to the translational entropy costs of assembly outweighing the energetic gain of assembly.  As the concentration of subunits is increased, above the critical aggregation concentration scale, the mass shifts to larger sized assemblies.  In both cases, where defects are stable or unstable, the primary peak in $\Phi_N$ shifts from monomers ($N=1$) to $N \simeq N_* \approx9$ continuously, which is a consequence of the lack of a nucleation barrier for 1D assemblies~\cite{Joswiak:2016,Chen:2018}~\footnote{Generic arguments for higher dimensional assemblies, imply a surface cost that grows faster than $N$ at small sizes (i.e. a nucleation barrier), and in general gives rise to two distinct peaks:  free monomers ($N \approx 1$) and primary aggregates ($N \approx N_*)$}.  When $-\delta_T>0$ as in case (v) and weak-binding is entropically unfavorable, the mass distribution stabilizes around the primary frustration limited aggregate size $N=N_*$ at high concentrations (Figure \ref{fig:example}b, right), which is a defining feature of {\it self-limiting assembly}.  In contrast, when $-\delta_T<0$ and weak-binding is comparably favorable to strong-binding(i.e $\delta_T \lesssim \Delta_T$) (e.g. case (ii)), defects are stable at high concentrations, leading to the mass distribution shifting to higher-$N$ populations corresponding to chains of primary $N_*$-mer aggregates held together by weak, defective bonds (Figure \ref{fig:example}b, left).  Hence, while this scenario corresponds to assemblies that are composed of {\it locally well-defined} sizes (i.e. frustration-limited chains of size $N_*$), the overall mass of aggregates is not well-defined  and essentially exhibits the concentration-dependent, exponential distribution characteristic of one-dimensional equilibrium chain assembly.  
  
		It is useful to establish some criteria for determining whether one has dispersed subunits, self-limiting assemblies, or unlimited and defective aggregates.  We show that this can be done by considering the mean aggregate size $\langle N \rangle$ and, in particular, the relative size fluctuations of the form $\frac{\langle\delta N^2\rangle}{\langle N^2\rangle}$ (Figures \ref{fig:example}c,d).  When weak binding is favorable as in case (i), the average size of assemblies $\langle N\rangle$ grows quite rapidly as expected.  However, as one lowers the weak binding free energy, a plateau in the average size begins to form.  We associate this slowdown in the growth of the average assembly size with self-limitation.  Note that while case (ii) does not strictly have local minima, the non-convexity of free energy near the self-limiting size is enough to stabilize a local peak in $\Phi_N$ near to $N_*$.  The range of concentrations over which the plateau in $\langle N\rangle$ persists increases as one reduces the weak binding free energy.  In particular for case (v), when weakly-bound structures are totally unfavorable, $\langle N\rangle$ remains stable up to a maximum concentration $\Phi_{\rm max}$.  For our purposes, we choose an upper limit to concentration $\Phi_{\rm max}=1$, since ideal-aggregation theory clearly breaks down in this regime, and it is not possible to have volume fractions that exceed (or even approach) unity.  
  
        Noting that self-limitation corresponds to a mean aggregate size that is {\it independent of concentration}, We can further quantify the concentration range over which self-limiting assembly occurs by computing the relative size fluctuations $\langle\delta N^2\rangle/\langle N^2\rangle$.  A useful relation is (see Appendix \ref{app:susceptibility and fluctuations})
		\begin{equation}
			\frac{\langle\delta N^2\rangle}{\langle N^2\rangle}=\frac{d(\log\langle N\rangle)}{d(\log\Phi)},
		\end{equation}
		that is, the relative size fluctuations is the same as the susceptibility of the average size to concentration changes.  Accordingly, as shown in Figure \ref{fig:example}d, there is a decrease in the fluctuations when the average size begins to plateau.  Based on this correspondence, we define the onset of self-limiting assembly as the concentration at which there is a local maximum in the fluctuations (black dotted line).  This estimates the concentration beyond which dispersed monomers are no longer the dominant structure in the system and larger structures begin to form. The growth of larger structures, particularly self-limiting ones, continues until the local minimum in the fluctuations~\footnote{A strictly self-limiting state would be well-described by a Gaussian peaked around some $N_* \geq 1$, and would have size fluctuations controlled by the convexity of the minimum in the free energy per subunit~\cite{Hagan:2021}.}.   This decrease in the fluctuations can be seen from Figure \ref{fig:example}b where $\langle\delta N^2\rangle$ of the mass distribution stabilizes while $\langle N^2\rangle$ increases.  We use this local minimum to define the onset of the formation of defective assemblies and a rise in size fluctuations.  As Figure \ref{fig:example}d shows, making weak binding unfavorable extends the range of self-limiting assembly.  In particular for case (v), there is no local minimum up to the maximum concentration, which indicates the suppression of larger defective structures.  It is useful to note that for $-\delta_T<0$ and sufficiently low temperatures or high concentrations, the energy landscape is nearly flat for large assembly sizes ($N \gg N_*$) and the distribution will have an exponential tail.  In those limits, the size fluctuations approach $\langle\delta N^2\rangle/\langle N^2\rangle\rightarrow1/2$, which is characteristic of equilibrium 1D chain assembly, and therefore, corresponds to the uncontrolled growth of defective aggregates.Hence, these cases ((i)-(iv) in Fig. \ref{fig:example}) exhibit {\it secondary aggregation behavior} according to the size-fluctuation dependent criteria, with a primary aggregation transition from free monomers to self-limiting $N_*$-mers at a lower critical concentration, followed by a secondary aggregation transition to the unlimited (defective) state of assembly at an upper critical concentration.  Notably, this self-limiting state at intermediate concentration occurs when the primary aggregate ($N=N_*$) is not the global minimum of the free energy per subunit.  In general, dominance of primary aggregates at intermediate concentration derives from the generically higher translation entropy per subunit in smaller $N$ structures that can compensate for the otherwise higher free energy to assemble those states than the larger $N$ groundstate.  Notably, the width of the intermediate concentration window of self-limiting assembly grows as the gap in energy between the $N_*$-mer and the unlimited, defective chain is reduced (i.e. as $\delta_T \to 0_+$)

         With these criteria for the onset of self-limiting assembly and the formation of defective assemblies defined, we turn to examining the role of temperature and frustration in self-limiting assembly in the next section.

	\section{\label{sec:phase diagrams}Role of temperature, concentration, and frustration}

        In this section, we now analyze the assembly behavior of the incommensurate polybrick model as a function of three key control parameters -- concentration $\Phi$, temperature $T$, and frustration $f$ -- focusing on parameters that control the stability of the self-limiting state relative to the disperse state and states of (defective and defect-free) unlimited assembly.
 
		\subsection{\label{subsec:T-Phi diagram}Temperature vs. concentration phase diagrams}		
			We start by fixing the physical parameters of subunits -- the frustration $f$, elastic and interaction parameters -- and consider how temperature $T$ and subunit concentration $\Phi$ influence the resulting assemblies.  In particular, we illustrate the case where frustration at $T=0$ selects a finite primary aggregate size is $N_*\approx9$.  As discussed in Sections \ref{subsec:bond entropies} and \ref{subsec:landscape}, there are two important temperatures: the temperature $T_{\delta}=\delta_0/s_{\delta}$ at which weakly-bound subunits melt and the temperature $T_{\Delta}=\Delta_0/s_{\Delta}$ at which strongly-bound subunits melt.  Thus, there are two limits of interest: $T_{\delta}<T_{\Delta}$ and $T_{\delta}>T_{\Delta}$.  To study these limits, we start by considering the case where the binding energy of defects are ``slightly sticky", $\delta_0/\Delta_0=0.3$.  We compare two cases of  ``conformational stiffness'' of the weak bonds (i.e. how much binding restricts the vibrational and orientational modes and reduced entropy):  relatively stiff defects, $s_{\delta}/s_{\Delta}=1.0$, where partial disruption of strong bonds does not significantly change the vibrational/orientational behavior subunits; and relatively floppy defects, $s_{\delta}/s_{\Delta}=0.1$, where weak bonds exhibit significantly enhanced conformational fluctuations over strong bonds.  These correspond to temperature ratios $T_{\delta}/T_{\Delta}=0.3$ and $T_{\delta}/T_{\Delta}=3.0$, respectively.

            In these cases, we consider the necessary conditions for self-limiting assembly, which in particular requires finite temperature fluctuations to destablize otherwise energetically favored weak binding.  In addition to the nominal free energy of weak binding, $-\delta_T =-\delta_0 +T s_\delta$, the thermodynamic stability of defective bonds will also depend on the translation entropy gain of breaking we bonds, which we denote as $s_{\rm trans} (\Phi)$ .  We expect a condition for the thermal stability of self-limiting assembly $T>T_{\rm min} (\Phi)$, defined by $-\delta_0 +T_{\rm min} \big[s_\delta+ s_{\rm trans} (\Phi) \big]=0$, or
            \begin{equation}
                T_{\min}(\Phi)=\frac{T_\delta}{1+s_{\rm trans} (\Phi) / s_\delta }.
                \label{eq: Tmin}
            \end{equation}
            A more careful treatment that considers the relative mass in (primary) self-limiting chains versus unlimited, defective assemblies (Appendix \ref{app:multiple minima approximation}) gives this same result with an approximate expression for the translation entropy gain of weak-bond breaking, $s_{\rm trans} (\Phi)=k_B \ln\frac{(2-\sqrt{2})N_{\star}}{\Phi}$.  
            This result shows that the thermal stability of self-limiting assembly occurs above a critical temperature that is shifted below the nominal melting temperature of defective bonds ($T_\delta$) by an amount that increases with concentration as well as conformational entropy cost of weak bonds.  We illustrate this effect on the thermal stability criterion for the regime of self-limitation below.

        	\begin{figure*}
				\centering
				\includegraphics[width=\textwidth]{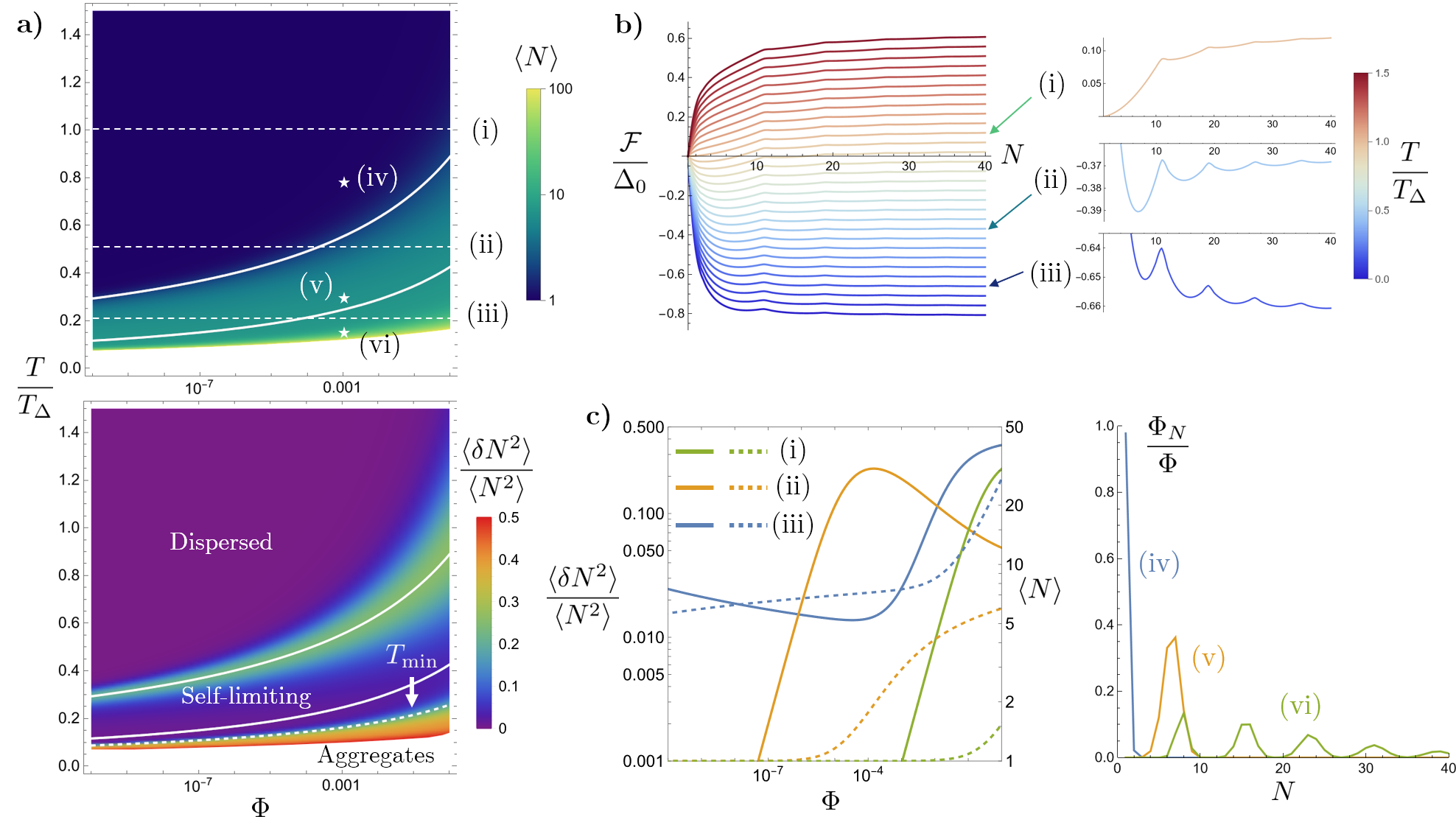}
				\caption{\textbf{a)} Concentration and temperature dependence of the average size and relative size fluctuations for $k_f/k_u=100$, $\delta_0/\Delta_0=0.3$, $k_ff^2a^2/2\Delta_0=1$, and $s_{\delta}/s_{\Delta}=1.0$ ($s_{\delta}=s_{\Delta}=10k_B$).  The ratio of melting temperatures is $T_{\delta}/T_{\Delta}=0.3$ ($k_B T_{\Delta}=0.1 ~ \Delta_0$).  Solid white lines are boundaries representing the onset of self-limiting assembly (upper white line) and the onset of the formation of defective aggregates (lower white line).  $T_{\rm min}(\Phi)$ is the minimum temperature below which defective aggregate dominate the system (dashed white line).  \textbf{b)} Free energy landscapes as a function of the temperature: (i) $T/T_{\Delta}=1.0$, (ii) $T/T_{\Delta}=0.5$, and (iii) $T/T_{\Delta}=0.2$.  In case (iii), $-\delta_T<0$ and defective aggregates are energetically favorable.  In case (ii), $-\delta_T>0$ and self-limiting assemblies are energetically/entropically favorable.  In case (i), $\Delta_T>0$ and dispersed subunits are favorable.  \textbf{c)} Average size (dashed) and relative size fluctuations (solid) for cases (i), (ii), and (iii), and the mass distribution of subunits for decreasing temperatures (iv) $T/T_{\Delta}=0.8$, (v) $T/T_{\Delta}=0.3$, and (vi) $T/T_{\Delta}=0.15$.}
				\label{fig:unfavorable defects}
			\end{figure*}
            \subsubsection{Stiff defects}
            We start with the $T_{\delta}/T_{\Delta}=0.3$ or the case of relatively stiff defective (weak) bonds.  The concentration-temperature phase diagram is shown in Figure \ref{fig:unfavorable defects}a.  The solid black lines are the boundaries determined from the size-fluctuation criteria discussed in Section \ref{sec:criteria} (i.e. where self-limitation corresponds to $\frac{d^2 (\log \langle N \rangle)}{d(\log\Phi)^2} <0$) that delineates states of dispersed monomers, self-limiting, and unlimited/defective assembly.  Due to the temperature dependence of binding free energies, the nature of the free energy subunit is depends strongly on $T$ (Figure \ref{fig:unfavorable defects}b), exhibiting variants of the fixed-$\Delta_T$ examples illustrated in Section \ref{sec:criteria}.  At low temperatures $T<T_{\delta}$ and favorable weak binding $-\delta_T<0$ (e.g. case (iii)), the defective aggregates states are thermodynamically stable.  This means that as the concentration of subunits increases, the larger defective states will eventually become occupied and the size of the assemblies will grow uncontrollably at large enough concentration (i.e. above a secondary critical aggregation concentration).  Similar to what was described in Section \ref{sec:criteria}, this can be seen by the rapid rise of the average assembly size $\langle N\rangle$ (Figure \ref{fig:unfavorable defects}c, dashed blue line) after it plateaus near the self-limiting size $N_{\star}\sim 9$, in addition to the local minimum in size fluctuations (solid blue line).  As the temperature is raised, the defective aggregates become less favorable until $-\delta_T>0$ as in case (ii).  As $\Phi$ increases, the system transitions from dispersed subunits to self-limiting assemblies but not to defective aggregates, at least up to the maximal concentration $\Phi=1$.  This can be seen in Figure \ref{fig:unfavorable defects}c where the average size (dashed orange curve) begins to plateau and the fluctuations (solid orange) decrease but do not reach a local minimum for $\Phi \leq 1$.  Finally, as temperature further increased, we eventually enter the regime where $\Delta_T\ge 0$ and we have the melting of strongly-bound subunits.  When this happens, the dominate state will obviously be dispersed subunits for any concentration.  This of course results in an average size that barely increases beyond a single subunit (dashed green curve) and fluctuations that do not reach a local maximum (solid green curve).  

            Also shown in Figure \ref{fig:unfavorable defects}a (bottom) as the dashed white line, is the prediction for $T_{\rm min}(\phi)$, which is generally in agreement with when the boundary defined by where the size fluctuations begin to plateau to 0.5, corresponding to the regime of uncontrolled growth of defective chains.
            
			\begin{figure*}
				\centering
				\includegraphics[width=\textwidth]{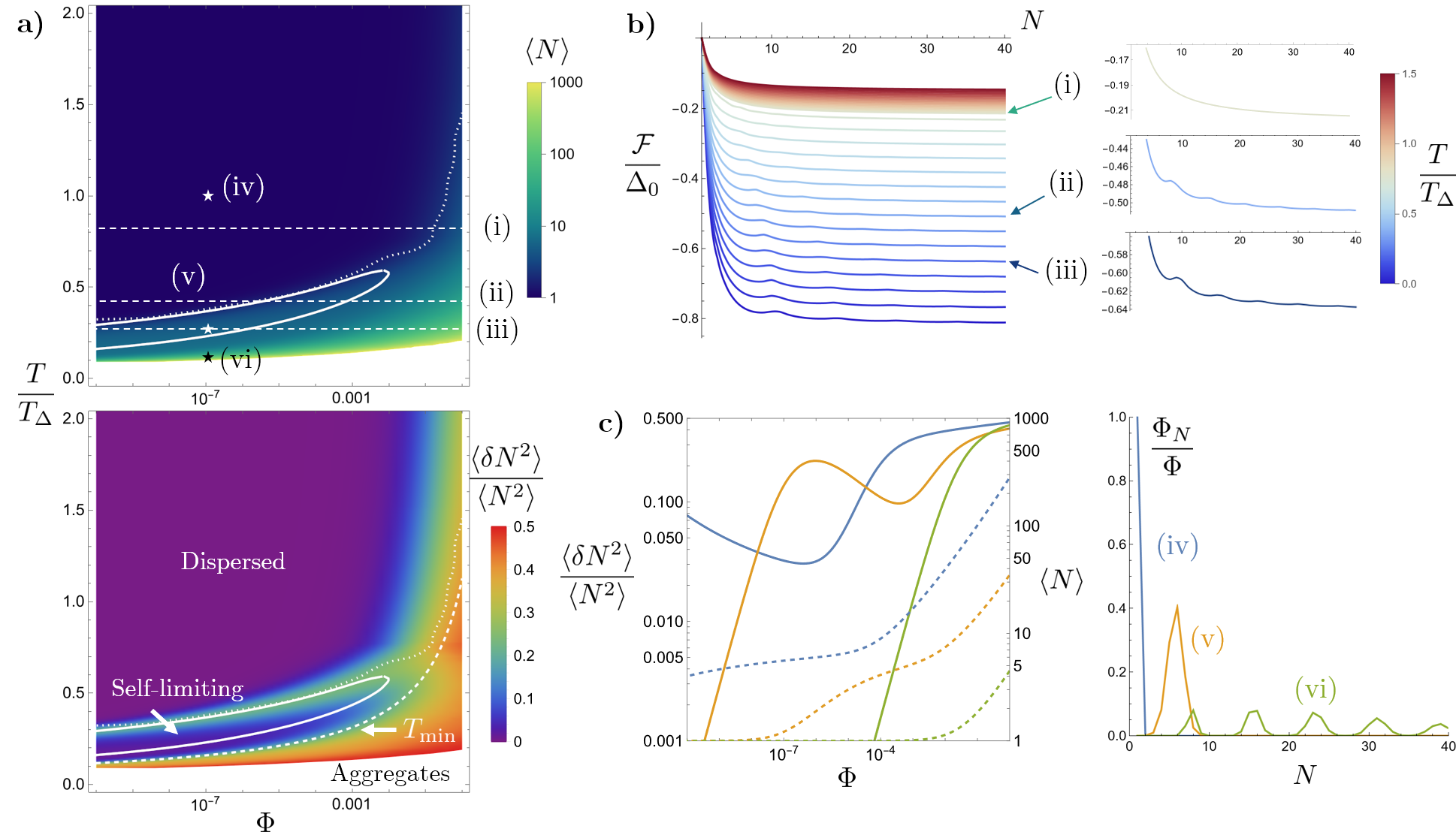}
				\caption{\textbf{a)} Concentration and temperature dependence of the average size and relative size fluctuations for $k_f/k_u=100$, $\delta_0/\Delta_0=0.3$, $k_ff^2a^2/2\Delta_0=1$, and $s_{\delta}/s_{\Delta}=0.1$ ($s_{\delta}=1 k_B $, $s_{\Delta}=10 k_B $).  The ratio of melting temperatures is $T_{\delta}/T_{\Delta}=3$ ($k_B T_{\Delta}=0.1 \Delta_0$).  Solid white lines are boundaries representing the onset of self-limiting assembly and the onset of the formation of defective aggregates.  The dotted white line indicates the transition out of dispersed subunits into either self-limiting assemblies or defective aggregates.  In this case, self-limiting assemblies do not exist at high concentrations.  $T_{\rm min}(\Phi)$ is the minimum temperature below which weakly-bound aggregates dominate the system (dashed white line).  \textbf{b)} Free energy landscapes as a function of the temperature: (i) $T/T_{\Delta}=0.8$, (ii) $T/T_{\Delta}=0.4$, and (iii) $T/T_{\Delta}=0.25$.  \textbf{c)} Average size (dashed) and relative size fluctuations (solid) for cases (i), (ii), and (iii), and the mass distribution of subunits for decreasing temperatures (iv) $T/T_{\Delta}=1.0$, (v) $T/T_{\Delta}=0.25$, and (vi) $T/T_{\Delta}=0.1$.}
				\label{fig:favorable defects}
			\end{figure*}
            \subsubsection{Floppy defects}
   
            We now turn to the opposite limit when $T_{\delta}/T_{\Delta}=3.0$ or the case of relatively floppy weak binding, whose concentration-temperature phase diagram is shown in Figure \ref{fig:favorable defects}a.  This case is strikingly different from the case of $T_{\delta}/T_{\Delta}<1$.  In particular, the self-limiting region terminates before reaching the maximum concentration.  As shown in Figure \ref{fig:favorable defects}b, due to $T_{\delta}/T_{\Delta}>1$, defective and unlimited assembly is stable relative to dispersed monomers at temperatures well above the nominal melting of strong bonds, an effect that derives from the enhanced conformational entropy exhibited by defective bonding.  This results in the system transitioning to the uncontrolled growth of defective aggregates at high concentrations or having a direct transition from dispersed subunits to defective aggregates.  As shown in Figure \ref{fig:favorable defects}c for case (i), the size fluctuations (green solid line) does not exhibit a local minimum but rather a simple rise, indicative of dispersed subunits, followed by a plateau towards 0.5, indicative of defective aggregates.  The direct ``transition'' between dispersed subunits to defective aggregates can be characterized by the knee of the curve (indicative of a peak in a higher derivative of $\langle N \rangle (\Phi)$), which can be computed by considering $\frac{d^2}{d(\log\Phi)^2}\left(\frac{\langle\delta N^2\rangle}{\langle N^2\rangle}\right)$ and locating the minimum or most negative value.  This offers an alternate criteria for determining when dispersed subunits begin to form larger structures, whether those structures be self-limiting or defect-ridden.  Indeed, as seen in Figure \ref{fig:favorable defects}, this criteria matches well with that of choosing the local maximum in the size fluctuations when there is a transition from dispersed subunits to self-limiting assemblies.

            
		\subsection{\label{subsec:T-f diagram}Temperature vs frustration phase diagram}
			\begin{figure*}
				\centering
				\includegraphics[width=\textwidth]{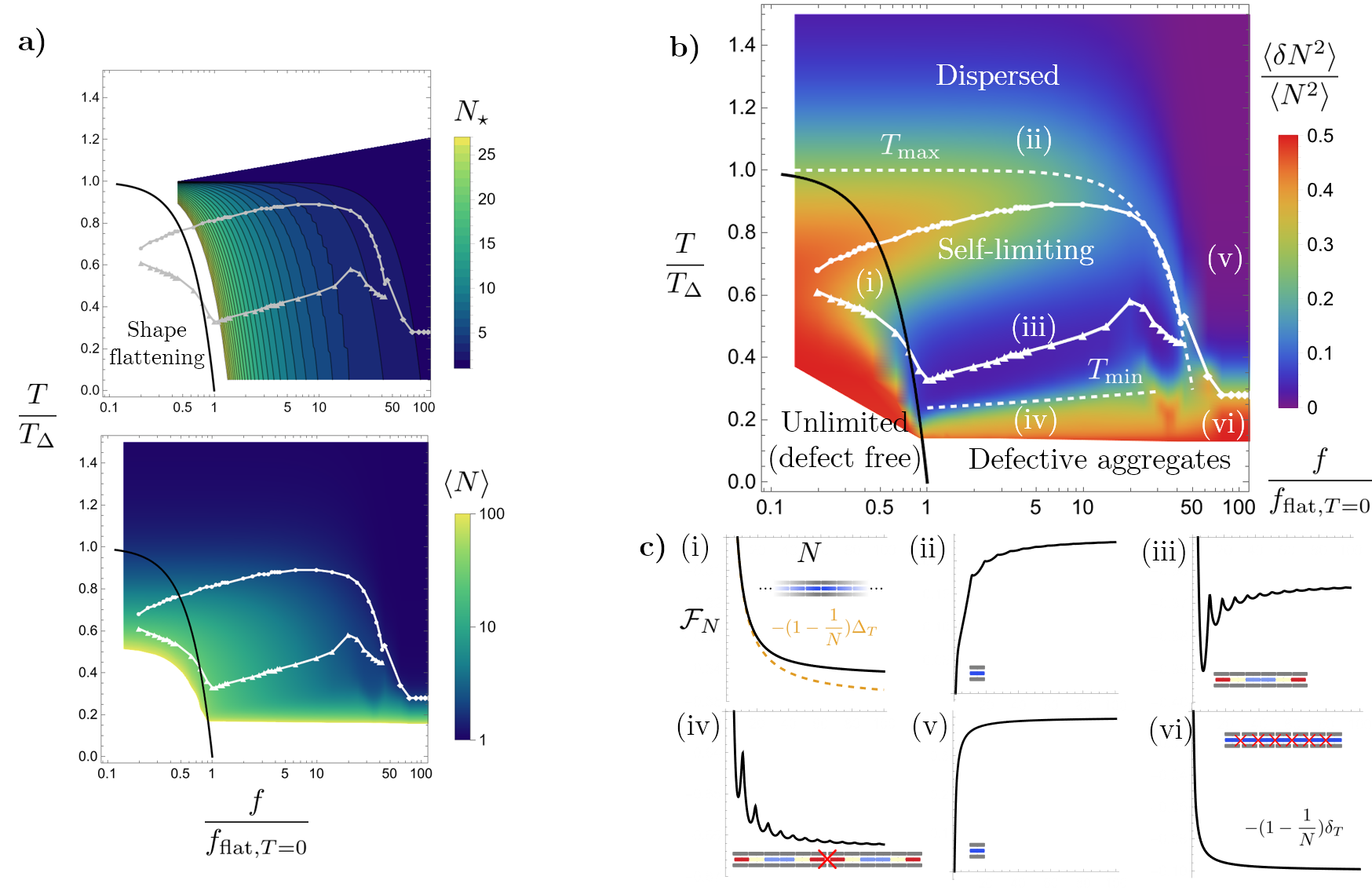}
				\caption{Frustration-temperature phase diagrams for $k_f/k_u=100$, $\delta_0/\Delta_0=0.3$, and $s_{\delta}/s_{\Delta}=1.0$ ($s_{\delta}=s_{\Delta}=10 k_B$).  The ratio of melting temperatures is $T_{\delta}/T_{\Delta}=0.3$ ($k_B T_{\Delta}=0.1 \Delta_0$).  \textbf{a)} Self-limiting size (top) and average assembly size (bottom) as functions of frustration and temperature.  \textbf{b)} Size fluctuations as functions of frustration and temperature.  The solid lines and points indicated the onsets for the formation of self-limiting assemblies and defective aggregates.  \textbf{c)} Free energy landscapes corresponding the various regimes indicated in b).}
				\label{fig:T vs f}
			\end{figure*}

            We now turn our attention to the role of frustration.  We consider the maximum concentration, in our case, $\Phi=1$ and focus on the types of assemblies that can form in the high concentration.  This focus is motivated by the attempt understand if a robust, self-limiting state is possible at sufficiently large, yet experimentally feasible concentration range.  We focus on the case of stiff weakly-bound subunits (Figure \ref{fig:unfavorable defects}) for which we expect there to be a finite-temperature window for self-limitation that can persist to high concentrations.  For each value of frustration, the criteria for determining the onset of self-limiting assembly and defect formation gives as an upper and lower temperature for self-limiting at the maximum concentration.  These boundaries are shown in Figure \ref{fig:T vs f} overlayed with the self-limiting size $N_{\star}$, the average assembly size $\langle N\rangle$, and the relative size fluctuations $\langle\delta N^2\rangle/\langle N^2\rangle$.  

            There are essentially three ranges of frustrations that are important.  When frustration is too small, the strain accumulation is not sufficient to prevent the binding of new subunits and assemblies undergo shape flattening, meaning that there is no local minimum at any finite $N$ and the assembly results in unlimited, strongly-bound and defect-free chains (as shown schematically in Figure \ref{fig:states}f) .  To be more precise, shape flattening occurs when (see Appendix \ref{subapp:continuum})
            \begin{equation}
                f\lesssim \sqrt{\frac{2\Delta_T}{k_fa^2\sqrt{4k_f/k_u}}}\equiv f_{\rm flat}(T).
                \label{eq:f flat}
            \end{equation}
            This is the finite-temperature generalization of the $T=0$ energetics considered in Section \ref{sec:model} above.   Note because $\Delta_T \to 0$ as temperature approaches $T_\Delta$, this critical threshold of frustration also vanishes, i.e. $f_{\rm flat}(T \to T_\Delta) \to 0$, and more generally decreases as temperature is raised. 
            
            In the opposite regime, When frustration is too large, it is frustration costs complete negate the energetic gains of subunit binding and so the self-limiting size approaches that of a single subunit.  This maximal value of frustration at which the self-limiting size $N_{\star}$ is roughly a single subunit is given by
            \begin{equation}
                f\gtrsim\sqrt{\frac{24\Delta_T}{k_ua^2}}\equiv f_{\rm single}(T).  
                \label{eq:f single}
            \end{equation}
            Therefore, self-limiting assembly most robustly occurs in the intermediate range of frustrations $f_{\rm flat}\lesssim f\lesssim f_{\rm single}$.  Note that the size of this range over which self-limiting assembly can occur is
            \begin{equation}
                \frac{f_{\rm single}}{f_{\rm flat}}\sim\left(\frac{k_f}{k_u}\right)^{\frac{3}{4}}.
            \end{equation}
            Recall (Section \ref{sec:model}) that $k_f/k_u$ represents the size scale over which strain can accumulate before the assembly flattens out.  Thus, increasing this size scale not only expands the regime of robust self-limiting assembly but also increases the size of such assemblies.

            Figure \ref{fig:T vs f}b-c summarize the high concentration phases and their corresponding example free energy landscapes.  The phase diagram can roughly be divided into six regions.  (i) At sufficiently low frustrations, assemblies undergo shape-flattening where there is a lack of a local minimum in the free energy landscape and growth is uncontrolled and unlimited, and takes the form of defect-free, strongly bound chains.  Note that the criteria for the onsets of self-limiting assembly and defect formation extend into the shape-flattening region.  This is due to the small amount of frustration slowing down the drop in free energy at large sizes.  (ii) At sufficiently high temperatures, it is entropically unfavorable for subunits to bind, and so the preferred state of the system is that of dispersed subunits.  (iii) As temperature is lowered but kept above $T_{\delta}$, the subunits begin to assembly into stable self-limiting structures since weakly-bound subunits are entropically unfavorable.  (iv) As temperature is further lowered below $T_{\delta}$, the defective aggregate states become energetically favorable.  As a result, defective aggregates always dominate at sufficiently low temperatures.  (v) At sufficiently high frustrations, it is too costly to fully bind subunits.  Above $T_{\delta}$, the favorable state is dispersed subunits.  Finally, (vi) below $T_{\delta}$, while fully bound subunits may be unfavorable due to high frustration, weakly-bound subunits can be energetically favorable due to stress attenuation from defect formation.  This will result in the assembly of linear chains made up of all weakly-bound subunits.  Note that the boundary separating the direct transition from dispersed subunits (v) to defective aggregates (vi) uses the same criteria we discussed in Section \ref{sec:phase diagrams} where the second derivative of the size fluctations can detect the onset of the plateau of the size fluctuation towards 0.5, which represents unlimited growth.  In fact, this boundary sits roughly at $T_{\delta}$ (in this case 0.3).

            In Appendix \ref{app:multiple minima approximation}, we derive the minimum and maximum temperatures between which self-limiting assemblies are robust.  The minimum temperature $T_{\rm min}(\Phi=1)$ below which a majority of subunits can be found in defective aggregates is
            \begin{equation}
                T_{\rm min}(\Phi=1)\simeq T_{\delta}\left[1+\frac{1}{s_{\delta}}\ln\left(\frac{24(2-\sqrt{2})^3\Delta_0}{k_uf^2a^2}\right)^{\frac{1}{3}}\right]^{-1},
                \label{eq:T min}
            \end{equation}
            which follows from eq. (\ref{eq: Tmin}) and the $T=0$ approximation for the primary, self-limiting size $N_{\star}\simeq \left(\Delta_0/k_uf^2a^2\right)^{1/3}$.  The maximum temperature $T_{\rm max}$ above which the system is dispersed (no strong or weak bonds despite being dense) is given by
            \begin{equation}
                T_{\max}(\Phi=1)\simeq  T_{\Delta}\left[1-\frac{9}{4(4k_f/k_u)^{3/2}}\left(\frac{f}{f_{\rm flat}}\right)^2\right],
                \label{eq:T max}
            \end{equation}
            which follows from considering the frustration-dependent excess energy cost of forming primary, self-limiting aggregates and the estimated melting point of strong (i.e. frustrated) binding.  Notably, these two equations show that the upper and lower temperature limits for self-limiting assembly are set, to a first approximation, by the nominal melting temperatures of strong and weak bonds, respectively, $T_\Delta$ and $T_\delta$.  The maximum frustration before the subunits are too frustrated to form fully-bound assemblies is $f_{\rm max}\sim f_{\rm flat}(k_f/k_u)^{3/4}\sim f_{\rm single}$.  Additionally, these results show that the temperature range of self-limitating assembly narrows with increasing $f$, due to both a slight increasing value of $T_{\rm min}(\Phi=1)$ with frustration, and more important, a strongly-decreasing dependence of $T_{\rm max}(\Phi=1)$ on $f$, which effective closes as $f \to f_{single}$.

            In Appendix \ref{app:varying weak-binding} and Figure \ref{fig:varying defect binding}, we illustrate these effects by comparing additional phase diagrams when the weak-binding $\delta_0/\Delta_0$ is varied, showing that the stable temperature window for self-limitation narrows as nominal melting of defects approaches that of the strong bonds (i.e. as $T_\delta \to T_\Delta)$.
            
	\section{\label{sec:discussion}Discussion and Conclusion}
        \begin{figure}
            \centering
            \includegraphics[width=\columnwidth]{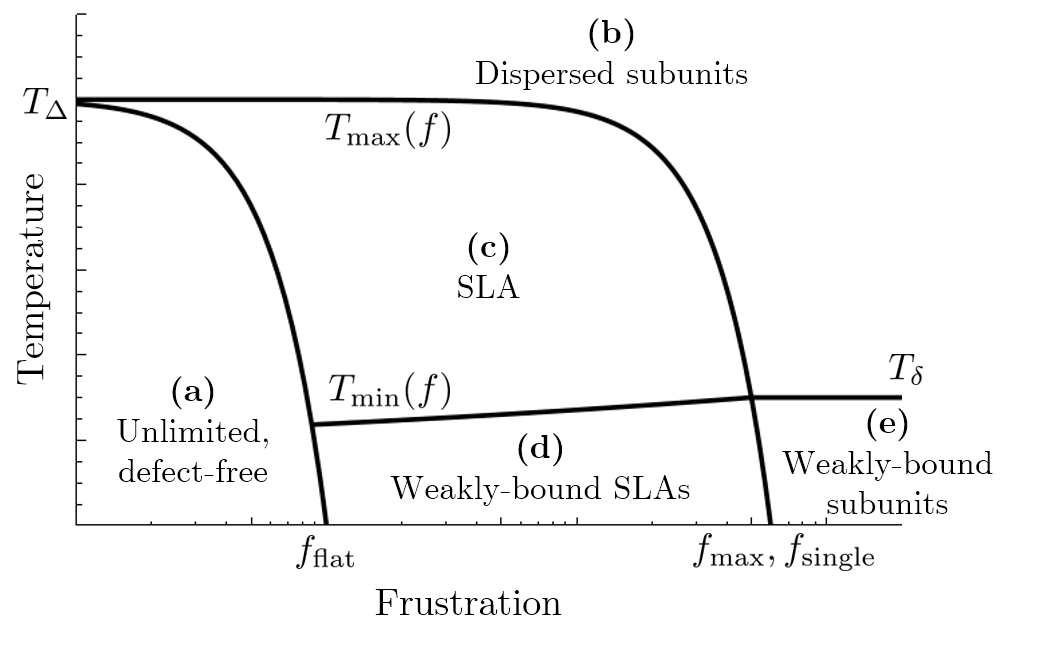}
            \caption{Schematic of frustration-temperature phase diagram of GFAs at high concentrations.  \textbf{a)} Low frustration and temperature: unlimited growth of shape-flattenend assemblies without any defects.  \textbf{b)} High frustration or temperature: dispersed subunits.  \textbf{c)} Intermediate frustration and temperature: self-limiting assemblies.  \textbf{d)} Intermediate frustration and low temperature: unlimited, defective aggregate of weakly-bound self-limiting assemblies.  Stable SLA can exist at low or intermediate concentrations.  \textbf{e)} High frustration and low temperature: unlimited, defective aggregate of weakly-bound subunits.}
            \label{fig:simplified phase diagram}
        \end{figure}
        In this article, we have introduced a solvable model of a frustrated incommensurate chain of ``polybricks'' to probe the role of temperature, concentration, and frustration (or subunit shape) in the process of self-limiting, geometrically frustrated assembly.  This model captures features of geometrically frustrated assemblies, particularly the competition between subunit binding and the accumulation of strains due to frustration as an assembly grows.  We focused on two categories of frustration escape: shape flattening, when frustration is insufficient to overcome subunit binding, and weak-binding, when assemblies can attenuate or stop stress propagation by breaking bonds or forming defects, which notably controls the thermal stabilty of self-limiting assembly.  The resulting free energy landscapes with both strong and weak binding allowed us to determine how temperature, concentration, and frustration control the stability of self-limiting assembly.

        Our main results are summarized in Figure \ref{fig:simplified phase diagram}.  The key result is that self-limiting assembly is robust at intermediate temperatures $T_{\delta}\lesssim T\lesssim T_{\Delta}$ set by the entropy costs of weak and strong binding, and frustrations $f_{\rm flat}\lesssim f\lesssim f_{\rm max}$ set by the elastic costs of binding.  At low frustration and temperature \textbf{(a)}, the strain accumulation from frustration is insufficient for overcoming the strong binding between subunits, so the assemblies effectively ignore frustration and grow unlimited.  The boundary for the shape-flattening regime is given by the frustration $f_{\rm flat}(T)$ (Eq.\ \ref{eq:f flat}), which is the minimum frustration needed to have self-limiting assembly.  At high temperatures or high frustrations (\textbf{b}), either the strong bonds are melted with $\Delta_T=\Delta_0-Ts_{\Delta}\ge0$ or $T\ge\Delta_0/s_{\Delta}=T_{\Delta}$, or the subunits are too frustrated to bind, resulting in dispersed subunits.  As the temperature is lowered below the weak-binding melting temperature $T_{\delta}=\delta_0/s_{\delta}$ at high frustrations (\textbf{e}), the weak binding free energy changes sign to $-\delta_T=-(\delta_0-Ts_{\delta})<0$, and so while strong binding is too costly due to being overly frustrated, the subunits can still weakly bind to form unlimited aggregates of weakly-bound subunits.  The boundary that describes the onset of dispersed subunits or unlimited aggregates of weakly-bound subunits is given by $T_{\rm max}(f)$ in Eq.\ (\ref{eq:T max}), from which the maximum frustration can be determined to be $f_{\rm max}\sim f_{\rm flat}(k_f/k_u)^{3/4}$, where $k_f/k_u$ is the ratio of inter- to intra-subunit stiffness.  Therefore, at intermediate frustrations $f_{\rm flat}\lesssim f\lesssim f_{\rm max}$ where we avoid shape-flattened assemblies and overly frustrated dispersed or aggregated subunits, we expect to obtain the target self-limiting assemblies (\textbf{c}).  The formation of self-limiting assemblies is most robust at high concentrations when the temperature is above $T_{\rm min}(f)$ (Eq.\ (\ref{eq:T min})), which is the temperature at which weak bonds entropically (including translational entropy) melt.  Below this temperature (\textbf{d}), defective aggregates of weakly-bound self-limiting assemblies become energetically favorable.  Note that while self-limiting assemblies in this regime of frustration and temperature are unstable at high concentrations, it is possible to stabilize them at lower concentrations.  It can be shown (see Appendix \ref{app:multiple minima approximation}) that the transition from dispersed to self-limiting and self-limiting to defective aggregation occurs at concentrations $\Phi_{\star}\sim e^{\beta\mathcal{F}_{\star}}$ and $\Phi_{\star\star}\sim e^{-\beta\delta_T}$, respectively, the ratio of which is then $\Phi_{\star\star}/\Phi_{\star}\sim e^{-\beta(\mathcal{F}_{\star}+\delta_T)}$.  The quantity $\mathcal{F}_{\star}+\delta_T$ can be interpreted as the difference between the free energies of a subunit in a self-limiting assembly and one in an aggregate with all weak bonds.  Thus if $\mathcal{F}_{\star}<-\delta_T$, it is possible to stabilize self-limiting assemblies over a low concentration range $\Phi_{\star}<\Phi<\Phi_{\star\star}<1$.  Note that as temperature is lowered, both $\Phi_{\star}$ and $\Phi_{\star\star}$ approach zero exponentially fast.  Thus for any concentration, defective aggregates will always dominate when the temperature is low enough.
        
        Our analysis offers some potential lessons in designing robust self-limiting assemblies.  
        \begin{itemize}
            \item[$\bullet$] The self-limiting size is controlled by the ratio of inter- to intra-subunit stiffnesses $\sqrt{k_f/k_u}$, which sets the size scale over which strains can accumulate before an assembly undergoes shape flattening.  The maximum self-limiting size can thus be increased by making inter-subunit interactions stiff while making the subunits themselves floppy.
            \item[$\bullet$] The range of frustrations $f_{\rm flat}\lesssim f\lesssim f_{\rm max}$ over which self-limiting assemblies can be stabilized, like the self-limiting size, is controlled by the quantity $k_f/k_u$ through the relation $f_{\rm max}/f_{\rm flat}\sim (k_f/k_u)^{3/4}$.  Thus, making inter-subunit interactions stiffer while making the subunits themselves floppiers has the additional benefit of allowing for a wider range of frustrations or subunit shapes.
            \item[$\bullet$] The minimum temperature needed to stabilize self-limiting assemblies $T_{\min}(f)$ (Eq.\ (\ref{eq: Tmin})) is primarily determined by the weak-binding melting temperature $T_{\delta}=\delta_0/s_{\delta}$, which accounts for the energy and entropy of weak-binding.  Lowering this temperatures requires either weakening the weak binding or increasing the entropy costs of weak binding.  When subunits are not strongly-bound, they should be as repulsive as possible or as restricted as possible in, for example, their orientational degrees of freedom.  In experiments \cite{Berengut:2020}, for instance, while poly-t extensions were used to change the frustration, they can be also thought of as creating some repulsion between the subunits if they are not fully bound.
        \end{itemize}
        We can estimate where experiments \cite{Berengut:2020} lie within our model and analysis.  In experiments, it was found that the central blocks thermally fluctuated distances of roughly $u\sim 2\textrm{ nm}$.  The resulting intra-subunit stiffness is roughly $k_u\sim\frac{k_BT}{u^2}\sim1\textrm{ pN/nm}$.  The inter-subunit interation resulted from the binding of multiple two base pair sites on the surface of the central blocks with binding energies of roughly $\Delta_0\sim60\textrm{ pN}\cdot\textrm{nm}$.  Given that base pairs have a size of roughly $0.34\textrm{ nm}$, we arrive at a stiffness of $k_f\sim100\textrm{ pN/nm}$, resulting in a stiffness ratio of $k_f/k_u\sim100$.  The length of each subunit is roughly $30\textrm{ nm}$ with frustrations of roughly $f\sim0.01-0.1$.  When subunits are (strongly or weakly) bound, the angle of rotational motion is roughly $\theta\sim1^{\circ}-10^{\circ}$, which results in an entropic contribution $s_{\Delta}\sim s_{\delta}\sim10 k_B$.  The estimated self-limiting size is $N_{\star}\sim3-15$, consistent with the assembly lengths observed in experiments.  The minimum and maximum frustrations are $f_{\rm flat}\sim0.01$ and $f_{\rm max}\sim1.0$.  The weak binding energy is difficult to estimate, so we assume it is roughly $\delta_0/\Delta_0\sim0.1-0.5$.  The melting temperatures are $T_{\Delta}\sim1-1.5 T$ and $T_{\delta}\sim0.1-0.7 T$, where $T$ here refers to room temperature.  It is thus reasonable to conclude that the experiments are likely within the self-limiting regime, but can also be tuned to dispersed and unlimited/defective assembly regimes with suitable adjustments of binding affinity.

        While our model and analysis captures some generic features of the process of self-limiting assembly, there are open questions.  In particular, an interesting question is what aspects of our analysis carry over to other examples of geometrically frustrated assemblies \cite{Uchida:2017,Serafin:2021,Tyukodi:2022,Tanjeem:2022,Spivack:2022,Hall:2023,Hall:2016,Hackney:2023} and other types of defects \cite{Meng:2014,Irvine:2010,Guerra:2018,Li:2019,Hall:2023,Hackney:2023,Bruss:2013}.  This includes understanding how the dimensionality of assemblies may affect the thermodynamics (e.g. there is a nucleation barrier in higher dimensions \cite{Joswiak:2016}) and defects such as disclinations and dislocations in these higher dimensional assemblies changes the strain accumulation and free energy landscapes.  Thus, while many aspects of GFA models are are non-universal, strain accumulation and superextensive energy growth as well as the existence of multiple free energy minima due to frustration escape are quite generic.  Hence, we expect that the qualitative behavior exhibited by the present 1D polybrick model, notably the existence of a minimal temperature of self-limiting assembly as well as secondary aggregation behavior, to be generic and key features of physical realizations more broadly.

    \section*{Acknowledgments}
        The authors are grateful to N. Hackney and M. Hagan for valuable discussions and input.  This work was supported by US National Science Foundation through award NSF DMR-2028885 and the Brandeis Center for Bioinspired SoftMaterials, an NSF MRSEC, DMR-2011846.  
 
	\appendix
	\section{\label{app:exact solution}Exact solution of linear chain}
        We here derive the main results for the excess energy and finite temperature corrections for the incommensurate chain model.
 
        \subsection{Discrete limit}
            The excess energy of a linear chain of $N$ subunits with central block displacements $\{u_n\}$ is
            \begin{equation}
                E_{\rm ex}=\sum_{n=1}^{N}\frac{1}{2}k_uu_n^2+\sum_{n=1}^{N-1}\frac{1}{2}k_f(u_{n+1}-u_n+fa)^2.
            \end{equation}
            In mechanical equilibrium, the equilibrium positions $\{u_n^{(\rm eq)}\}$ satisfy
            \begin{equation}
                \sum_{n=1}^NK_{mn}u_n^{(\rm eq)}=k_ffa(\delta_{m,1}-\delta_{m,N}),
                \label{eq:mechanical eq discrete}
            \end{equation}
            where the tridiagonal matrix of spring constants $K_{mn}$ is given by
            \begin{equation}
                \boldsymbol{K}=\bmat{k_u+k_f & -k_f & \dots & 0 & 0 \\ -k_f & k_u+2k_f & \dots & 0 & 0 \\ \vdots & \vdots & \ddots & \vdots & \vdots \\ 0 & 0 & \dots & k_u+2k_f & -k_f \\ 0 & 0 & \dots & -k_f & k_u+k_f}{1.0}.
                \label{eq:k matrix}
            \end{equation}
            Using the equation for mechanical equilibrium (Eq.\ (\ref{eq:mechanical eq discrete})), we can rewrite the ground state excess energy in terms of the equilibrium displacements as
            \begin{equation}
                E_{\rm ex}[\{u_n^{(\rm eq)}\}]=\frac{1}{2}k_ffa\left[u_N^{(\rm eq)}-u_1^{(\rm eq)}\right]+\frac{N-1}{2}k_ff^2a^2.
            \end{equation}
            So we only need to determine the positions of the end subunits $u_1^{(\rm eq)}$ and $u_N^{(\rm eq)}$.  By symmetry, we have $u_1^{(\rm eq)}=-u_N^{(\rm eq)}$.  Inverting Eq.\ \ref{eq:mechanical eq discrete}, we have
            \begin{equation}
                u_N^{(\rm eq)}=k_ffa\left(K_{N,1}^{-1}-K_{N.N}^{-1}\right).
            \end{equation}
            Using properties of certain tridiagonal matrices \cite{Yueh:2008}, we have
            \begin{subequations}
                \begin{align}
                    K_{N,1}^{-1}&=\frac{k_f^{N-1}}{\det(\boldsymbol{K})}\\
                    K_{N,N}^{-1}&=\frac{1}{\det(\boldsymbol{K})}\prod_{n=1}^{N-1}\left[k_u+4k_f\sin^2\frac{(2n-1)\pi}{2(2N-1)}\right],
                \end{align}
            \end{subequations}
            where
            \begin{equation}
                \det(\boldsymbol{K})=\prod_{n=1}^N\left[k_u+4k_f\sin^2\frac{(n-1)\pi}{2N}\right].
                \label{eq:det K}
            \end{equation}
            Using these results, the excess energy can be written as
            \begin{equation}
                \mathcal{E}_{\rm ex}=\frac{k_ff^2a^2}{N}\left[\frac{N-1}{2}+k_f\left(K_{N,1}^{-1}-K_{N,N}^{-1}\right)\right].
            \end{equation}

		\subsection{\label{subapp:continuum}Continuum limit}
            In the continuum limit, which can be obtained by taking $\frac{u_{n+1}-u_n}{a}\approx\frac{du}{dx}$ and approximating the sums as integrals, the excess energy becomes
            \begin{equation}
                E_{\rm ex}=\int_{-\frac{L}{2}}^{\frac{L}{2}}dx\left[\frac{1}{2}Y_uu^2+\frac{1}{2}Y_fa^2\left(\frac{du}{dx}+f\right)^2\right],
            \end{equation}
            where $Y_u=k_u/a$ and $Y_f=k_f/a$ are the elasticities per unit length.  Minimizing with respect to the displacement field $u(x)$ yields
            \begin{equation}
                \frac{d^2u}{dx^2}=\frac{1}{\lambda^2}u,
            \end{equation}
            where $\lambda=\sqrt{Y_fa^2/Y_u}=a\sqrt{k_f/k_u}$ is the length scale over which strains accumulate, and the free boundary condition
            \begin{equation}
                \frac{du(\pm L/2)}{dx}=-f.
            \end{equation}
            The solution for the displacement field is
            \begin{equation}
                u(x)=-f\lambda\frac{\sinh x/\lambda}{\cosh L/2\lambda},
            \end{equation}
            and the excess energy density due to stain accumulation is
            \begin{equation}
                \mathcal{E}_{\rm ex}=\frac{1}{2}Y_ff^2a^2\left(1-\frac{\tanh L/2\lambda}{L/2\lambda}\right).
            \end{equation}
        
		\subsection{\label{subapp:entropy}Vibrational and orientational entropies}
\begin{figure}
                \centering
                \includegraphics[width=\columnwidth]{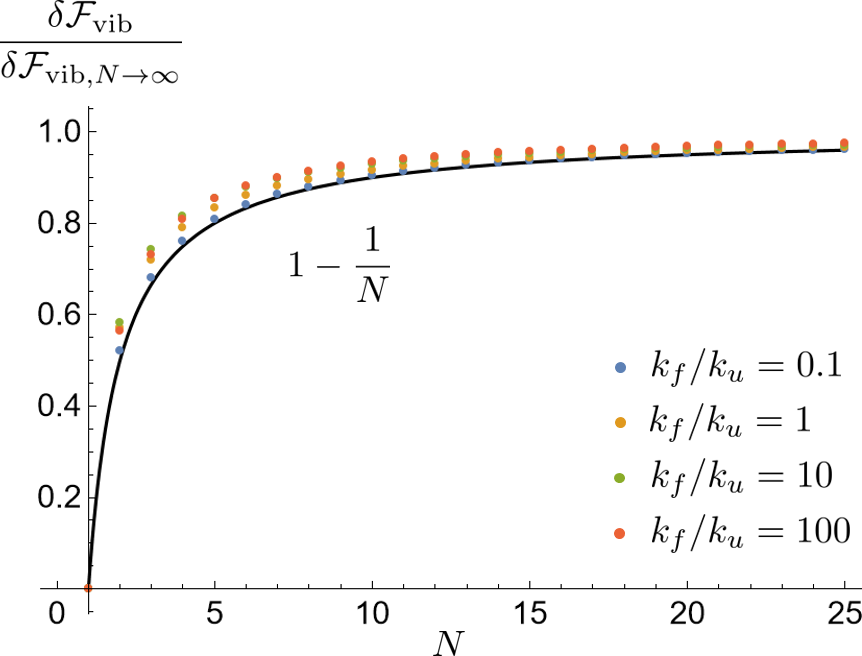}
                \caption{$1-\frac{1}{N}$ approximation (solid black line) for the finite-temperature vibrational contribution to the free energy (Eq. (\ref{eq:Fvib})).}
                \label{fig:Fvib approx}
            \end{figure}

            The finite-temperature orientational contributions to the free energy can be computed by assuming that while a free subunit can orient anywhere on the unit sphere, a bound subunit is restricted to a cone with angle $\theta_{\Delta}$.  The partition function is
            \begin{equation}
                Z_{\rm orient}=\left(\int_0^{2\pi}\int_0^{\theta_{\Delta}}\frac{d\Omega}{4\pi}\right)^{N-1}=\left(\frac{1-\cos\theta_{\Delta}}{2}\right)^{N-1}.
            \end{equation}
            The vibrational contribution can be obtained from the partition function by summing over all displacements $\{u_n\}$.  By considering displacements $\{\delta u_n\}$ about the equilibrium positions $\{u_n^{(\rm eq)}\}$, we can write the partition function as
            \begin{align}
                \begin{split}
                    Z_{\rm vib}&=\int\left(\prod_{n=1}^{N}d\delta u_n\right)e^{-\beta E_N[\{u_n^{(\rm eq)}+\delta u_n\}]]}\\
                    &=e^{-\beta E_N[\{u_n^{(\rm eq)}\}]}\int\left(\prod_{n=1}^{N}d\delta u_n\right)e^{-\beta\frac{1}{2}\delta u_mK_{mn}\delta u_n}\\
                    &=e^{-\beta E_N[\{u_n^{(\rm eq)}\}]}\sqrt{\frac{(2\pi)^N}{\beta^N\det(\boldsymbol{K})}},
                \end{split}
            \end{align}
            where $K_{mn}$ is the matrix of spring constants given by Eq.\ (\ref{eq:k matrix}) and $E_N[\{u_n^{(\rm eq)}\}]$ is the ground state energy.  The determinant if given by Eq.\ \ref{eq:det K}.  Taking the logarithm of these partition functions and shifting the free energy so that $F_{N=1}=0$, we arrive at the total free energy per subunit
            \begin{equation}
                \mathcal{F}=\mathcal{E}_N+\delta\mathcal{F}_{\rm vib}+\delta\mathcal{F}_{\rm orient},
            \end{equation}
            where
            \begin{subequations}
                \begin{align}
                    \delta\mathcal{F}_{\rm orient}&=\left(1-\frac{1}{N}\right)k_B T\ln\frac{2}{1-\cos\theta_{\Delta}},\\
                    \delta\mathcal{F}_{\rm vib}&=\frac{k_B T}{2N}\sum_{n=1}^{N}\ln\left[1+\frac{4k_f}{k_u}\sin^2\frac{(n-1)\pi}{2N}\right].
                    \label{eq:Fvib}
                \end{align}
            \end{subequations}
            In the limit of large assemblies ($N\rightarrow\infty$), the vibrational free energy becomes
            \begin{align}
                \begin{split}
                    \lim_{N\rightarrow\infty}\delta\mathcal{F}_{\rm vib}&=k_B T\int_0^{\frac{1}{2}}dz\ln\left(1+\frac{4k_f}{k_u}\sin^2\pi z\right)\\
                    &=k_B T\ln\left[\frac{1}{2}\left(1+\sqrt{1+\frac{4k_f}{k_u}}\right)\right].
                \end{split}
            \end{align}
            A useful approximation for our purposes is
            \begin{equation}
                \delta\mathcal{F}_{\rm vib}\approx\left(1-\frac{1}{N}\right)k_B T\ln\left[\frac{1}{2}\left(1+\sqrt{1+\frac{4k_f}{k_u}}\right)\right].
            \end{equation}
            This approximation is illustrated in Figure \ref{fig:Fvib approx}.

            For weak binding, we can compute, for example, the orientational free energy as follows.  Given an assembly of length $N$ that is made of $M$ weakly-bound structures, there are $N-M$ strong bonds and $M-1$ weak bonds.  Assuming that strong bonds and weak bonds can rotate about angles $\theta_{\Delta}$ and $\theta_{\delta}$, respectively, we can write the partition function as
            \begin{equation}
                Z_{\rm orient}=\left(\int_0^{2\pi}\int_0^{\theta_{\Delta}}\frac{d\Omega}{4\pi}\right)^{N-M}\left(\int_0^{2\pi}\int_0^{\theta_{\delta}}\frac{d\Omega}{4\pi}\right)^{M-1}.
            \end{equation}
            The free energy is then
            \begin{align}
                \begin{split}
                    \delta\mathcal{F}_{\rm vib}=\ &\frac{(N-M)k_B T}{N}\ln\frac{2}{1-\cos\theta_{\Delta}}\\
                    &+\frac{(M-1)k_B T}{N}\ln\frac{2}{1-\cos\theta_{\delta}}.
                \end{split}
            \end{align}
            The first term can be combined with the $N-M$ strong bonds, resulting in the temperature-corrected strong binding energy $\Delta_0-Ts_{\Delta}$, while the second term can be combined with the $M-1$ weak bonds, resulting in $\delta_0-Ts_{\delta}$, where for the case of the orientational contribution
            \begin{equation}
                s_{\delta}=k_B \ln\frac{2}{1-\cos\theta_{\delta}}.
            \end{equation}
		
	\section{\label{app:multiple minima approximation}Multiple minima approximation}
        As discussed in Section \ref{subsec:landscape}, the free energy landscape of the incommensurate chain with weak binding can have multiple local minima associated with the weak binding of multiple self-limiting assemblies.  We here describe a useful approximation where we assume that the subunits in the system are either dispersed, in self-limiting assemblies of size $N_{\star}$, or in defective weakly-bound aggregates of self-limiting assemblies of size $mN_{\star}$ (for $m>1$).  The free energy of a structure made up of $m$ weakly-bound self-limiting assemblies is
        \begin{equation}
            \mathcal{F}_{mN_{\star}}=\mathcal{F}_{\star}-\frac{(m-1)\delta_T}{mN_{\star}}.
        \end{equation}
        The determine $N_{\star}$ and $\mathcal{F}_{\star}$, note that in the strain accumulation regime ($N\lesssim\sqrt{4k_f/k_u}$) the free energy per subunit is
        \begin{equation}
            \mathcal{F}_N\simeq-\left(1-\frac{1}{N}\right)\Delta_T+\frac{1}{6}k_ff^2a^2\left(\frac{N}{\sqrt{4k_f/k_u}}\right)^2.
        \end{equation}
        Minimizing with respect to $N$, we find the self-limiting size
        \begin{equation}
            \frac{N_{\star}}{\sqrt{4k_f/k_u}}\simeq\left(\frac{3\Delta_T}{k_ff^2a^2\sqrt{4k_f/k_u}}\right)^{\frac{1}{3}}.
        \end{equation}
        Substituting this in to the free energy, we have
        \begin{equation}
            \mathcal{F}_{\star}\simeq-\Delta_T+\frac{1}{2}k_ff^2a^2\left(\frac{3\Delta_T}{k_ff^2a^2\sqrt{4k_f/k_u}}\right)^{\frac{2}{3}}.
            \label{eq:F star approx}
        \end{equation}
        
        Applying ideal aggregation theory to a system of dispersed subunits and assemblies of size $mN_{\star}$, we have for the total subunit volume fraction
        \begin{align}
            \begin{split}
                \Phi&=\Phi_1+\sum_{m=1}^{\infty}mN_{\star}\left[\Phi_1e^{-\beta\mathcal{F}_{\star}}e^{\beta\frac{(m-1)\delta_T}{mN_{\star}}}\right]^{mN_{\star}}\\
                &=\Phi_1+\frac{N_{\star}\left(\Phi_1e^{-\beta\mathcal{F}_{\star}}\right)^{N_{\star}}}{\left[1-\left(\Phi_1e^{-\beta\mathcal{F_{\star}}}\right)^{N_{\star}}e^{\beta\delta_T}\right]^2}.
            \end{split}
        \end{align}
        The second term can be split into the volume fractions of subunits in self-limiting structures $\Phi_{\rm sl}(\Phi_1)=N_{\star}\left(\Phi_1e^{-\beta\mathcal{F}_{\star}}\right)^{N_{\star}}$ and defective aggregates $\Phi_{\rm agg}(\Phi_1)=\Phi(\Phi_1)-\Phi_1-\Phi_{\rm sl}(\Phi_1)$.

        We can use this approximation to determine the conditions under which dispersed subunits transition to self-limiting assemblies and self-limiting assemblies transition to defective aggregates.  For the transition between dispersed subunits to self-limiting assemblies, note that the concentration scale $\Phi_1^{\star}$ at which $\Phi_1^{\star}=\Phi_{\rm sl}(\Phi_1^{\star})$ is given by
        \begin{equation}
            \Phi_1^{\star}=\left(\frac{e^{N_{\star}\beta\mathcal{F}_{\star}}}{N_{\star}}\right)^{\frac{1}{N_{\star}-1}}.
        \end{equation}
        Assuming that the mass of defective structures is negligible at this concentration scale, we can write the total concentration of subunits as $\Phi=2\Phi_1^{\star}$.  Taking the maximum concentration $\Phi_{\rm tot}=1$ and rearranging, we arrive at
        \begin{equation}
            e^{\beta\mathcal{F}_{\star}}=\left(\frac{N_{\star}}{2^{N_{\star}-1}}\right)^{\frac{1}{N_{\star}}}.
        \end{equation}
        Note that the righthand side weakly depends on $N_{\star}$ and is bounded between $1/2$ and $1$.  Therefore, as an estimate, the transition from dispersed subunits to self-limiting assemblies occurs roughly when $\mathcal{F}_{\star}$ changes sign.  This corresponds to when the self-limiting assemblies between energetically favorable over the dispersed subunits.  From Eq.\ (\ref{eq:F star approx}), we have the relation between temperature $T$ and frustration $f$
        \begin{equation}
            \frac{T}{T_{\Delta}}\approx1-\frac{9}{4(4k_f/k_u)^{3/2}}\left(\frac{f}{f_{\rm flat}}\right)^2.
        \end{equation}
        This temperature is the maximum temperature $T_{\rm max}$ above which the system will remain dispersed (see Section \ref{subsec:T-f diagram}).
        
        For the transition between self-limiting assemblies and defective aggregates, we note that the concentration scale $\Phi_1^{\star\star}$ at which $\Phi_{\rm sl}(\Phi_1^{\star\star})=\Phi_{\rm agg}(\Phi_1^{\star\star})$ satisfies
        \begin{equation}
            \left(\Phi_1^{\star\star}e^{-\beta\mathcal{F}_{\star}}\right)^{N_{\star}}e^{\beta\delta_T}=1-\frac{1}{\sqrt{2}}.
        \end{equation}
        This can be substituted back into the total volume fraction to obtain the equation of state
        \begin{equation}
            \Phi=(2-\sqrt{2})N_{\star}(f,T)e^{-\beta(\delta_0-Ts_{\delta})},
        \end{equation}
        which relates the temperature $T$, frustration $f$, and concentration $\Phi$ at which defective aggregates dominate over self-limiting assemblies.  For simplicity, if we assume that the self-limiting size $N_{\star}$ is fixed or does not vary significantly over some range of temperatures and frustrations, we can solve for the minimum temperature above which self-limiting assembly occurs.  This temperature $T_{\rm min}$ is given by
        \begin{align}
            \begin{split}
                T_{\rm min}&=\delta_0\left[s_{\delta}+k_B \ln\frac{(2-\sqrt{2})N_{\star}}{\Phi}\right]^{-1}\\
                &=\frac{T_{\delta}}{1+s_{\rm trans}(\Phi)/s_{\delta}},
            \end{split}
        \end{align}
        where $s_{\rm trans}$ captures the effect of translational entropy.
	
	\section{\label{app:susceptibility and fluctuations}Relation between average size and relative size fluctuations}
        The distribution of assemblies of size $N$ is defined as $\rho_N=\Phi_N/N=\left(\Phi_1e^{-\beta\mathcal{F}_N}\right)^N$.  The average size is therefore 
        \begin{equation}
            \langle N\rangle=\frac{\sum_{N=1}^{\infty}N\rho_N}{\sum_{N=1}^{\infty}\rho_N}=\frac{\sum_{N=1}^{\infty}N\left(\Phi_1e^{-\beta\mathcal{F}_N}\right)^N}{\sum_{N=1}^{\infty}\left(\Phi_1e^{-\beta\mathcal{F}_N}\right)^N}.
        \end{equation}
        Using $d\rho_N/d\Phi_1=N\rho_N\Phi_1^{-1}$, we have
        \begin{align}
            \begin{split}
                \frac{d\langle N\rangle}{d\Phi_1}&=\Phi_1^{-1}\frac{\sum_{N=1}^{\infty}N^2\rho_N}{\sum_{N=1}^{\infty}\rho_N}-\Phi_1^{-1}\left(\frac{\sum_{N=1}^{\infty}N\rho_N}{\sum_{N=1}^{\infty}\rho_N}\right)^2\\
                &=\Phi_1^{-1}\left(\langle N^2\rangle-\langle N\rangle^2\right),
            \end{split}
        \end{align}
        and
        \begin{equation}
            \frac{d\Phi}{d\Phi_1}=\frac{d}{d\Phi_1}\sum_{N=1}^{\infty}N\rho_N=\Phi_1^{-1}\langle N^2\rangle\sum_{N=1}^{\infty}\rho_N=\frac{\langle N^2\rangle\Phi}{\langle N\rangle\Phi_1}.
        \end{equation}
        Therefore
        \begin{equation}
            \frac{d\langle N\rangle}{d\Phi}=\frac{d\langle N\rangle}{d\Phi_1}\frac{d\Phi_1}{d\Phi}=\frac{\langle\delta N^2\rangle}{\langle N^2\rangle}\frac{\langle N\rangle}{\Phi_1},
        \end{equation}
        which can be rewritten as
        \begin{equation}
            \frac{d(\log\langle N\rangle)}{d(\log\Phi)}=\frac{\langle\delta N^2\rangle}{\langle N^2\rangle}.
        \end{equation}
        It is useful to note that for an exponential distribution $\langle\delta N^2\rangle/\langle N^2\rangle\rightarrow1/2$, which in our case is what happens at large sizes.
	
	\section{\label{app:varying weak-binding}Effect of weak-binding energy on the transition between self-limiting assemblies and defective aggregates}
        \begin{figure*}
			\centering
			\includegraphics[width=\textwidth]{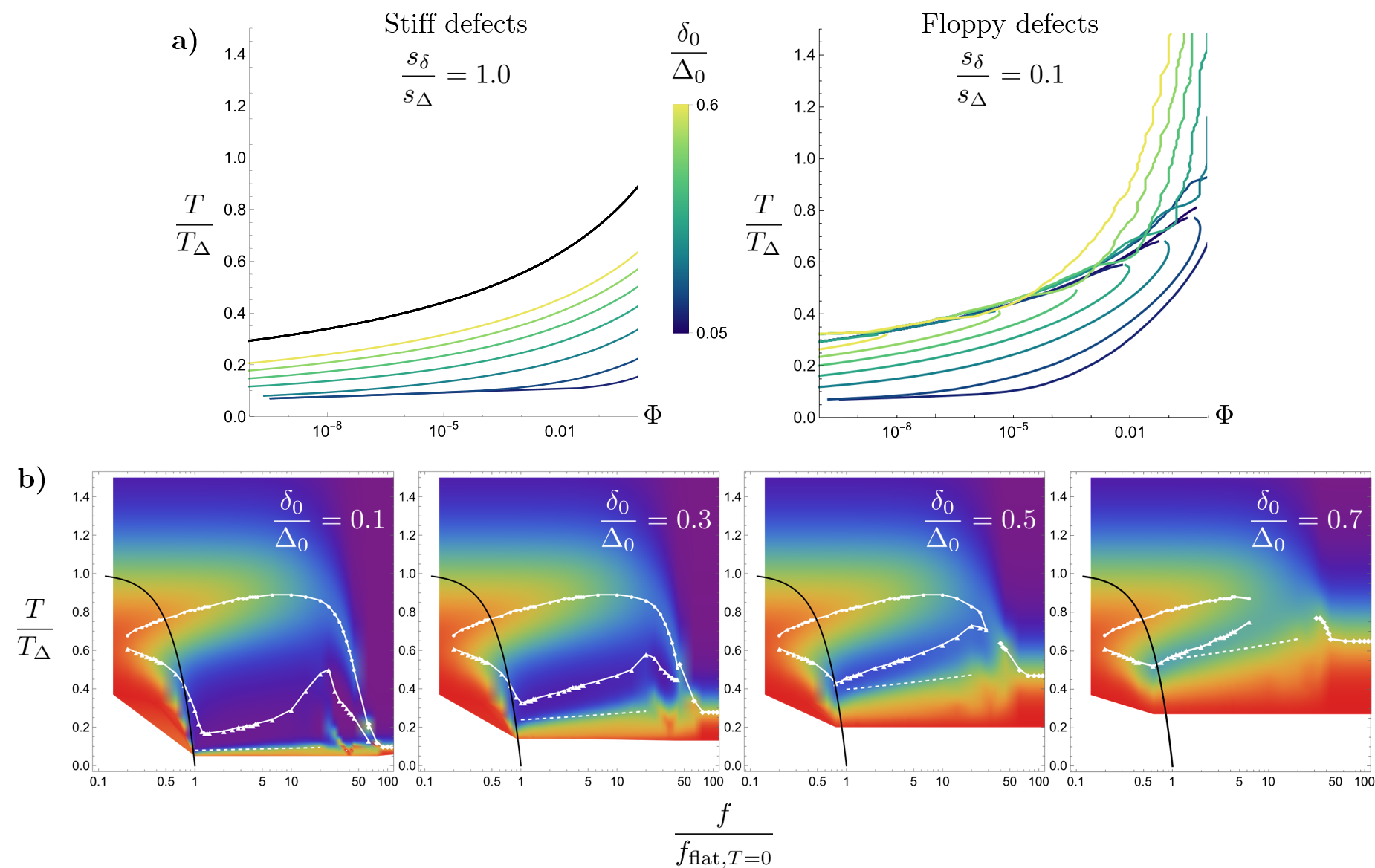}
			\caption{\textbf{a)} Boundaries for the onsets of self-limiting assembly and weakly-bound aggregation for varying weak-binding energy with $k_f/k_u=100$, $s_{\Delta}=10$, $k_ff^2a^2/2\Delta_0=1$, and $T_{\Delta}=0.1$.  \textbf{b)} Frustration-temperature phase diagrams for varying weak-binding energy for stiff defects ($s_{\delta}/s_{\Delta}=1$).  Increasing the weak-binding energy $\delta_0$ shifts the onset of defective aggregate formation (lower boundary) towards higher temperatures, reducing the regime of self-limiting assembly at high concentrations.}
			\label{fig:varying defect binding}
		\end{figure*}
        Figure \ref{fig:varying defect binding} shows how varying the weak-binding energy $\delta_0$ affects the onset of the transition from self-limiting assemblies to defective aggregates.  As the weak-binding energy is increased and weak binding becomes more energetically favorable, the self-limiting regime becomes more and more narrow.

\end{document}